\documentclass[epj]{svjour}
\usepackage{mathptmx}

\usepackage{amssymb}
\usepackage{amsmath}
\usepackage{cite}

\usepackage{graphicx}
\usepackage{epstopdf}
\DeclareGraphicsExtensions{.eps}

\begin{document}

\sloppy

\title{
 	Node-weighted measures for complex networks 
 	with spatially embedded, sampled, or differently sized nodes 
}

\date{Received: \today}


\author{
	Jobst Heitzig\inst{1} 
	\and 
	Jonathan F. Donges\inst{1, 2}
	\and 
	Yong Zou\inst{1, 3}
	\and 
	Norbert Marwan\inst{1}
	\and 
	J\"urgen Kurths\inst{1, 2, 4}
}
\institute{%
	Potsdam Institute for Climate Impact Research, Transdisciplinary Concepts and Methods, P.\,O.\ Box 60 12 03, 14412 Potsdam, Germany, \email{heitzig@pik-potsdam.de} 
	\and%
	Department of Physics, Humboldt University Berlin, Newtonstr.~15, 12489 Berlin, Germany
	\and%
    Department of Electronic and Information Engineering, Hong Kong Polytechnic University, Hung Hom, Kowloon, Hong Kong
	\and%
	Institute for Complex Systems and Mathematical Biology, University of Aberdeen,
	Aberdeen AB 24 UE, United Kingdom 
	}

\abstract{
  When network and graph theory are used in the study of complex systems, a
  typically finite set of nodes of the network under consideration is frequently
  either explicitly or implicitly considered representative of a much larger
  finite or infinite region or set of objects of interest. The selection procedure,
  e.\,g., formation of a subset or some kind of discretization or aggregation,
  typically results in individual nodes of the studied network representing
  quite differently sized parts of the domain of interest. This heterogeneity
  may induce substantial bias and artifacts in derived network statistics. To
  avoid this bias, we propose an axiomatic scheme based on the idea of {\em node
  splitting invariance} to derive consistently weighted variants of various
  commonly used statistical network measures. The practical relevance and
  applicability of our approach is demonstrated for a number of example networks
  from different fields of research, and is shown to be of fundamental
  importance in particular in the study of spatially embedded functional
  networks derived from time series as studied in, e.\,g., neuroscience and climatology. 
}

%
%

\maketitle

\def\nsi{n.\,s.\,i.}
\def\Nsi{N.\,s.\,i.}

\def\nodes{{\cal N}}
\def\edges{{\cal E}}
\def\cell{{\cal R}}

\section{\label{sec:intro}%
  Introduction
}

\subsection{Motivation}

\noindent In the last decades, network and graph theory have successfully been
applied to various kinds of complex systems, and many different measures have
been defined to study their structural and topological properties. Most of these
are of a combinatorial nature, based on counts of certain nodes, links,
triangles, paths, etc.\ (for an overview, see, e.\,g.,
\cite{Newman2003,Boccaletti2006,DaFCosta2007,Newmanbook2010,Cohenbook2010,Kolaczykbook2010}).

\begin{figure*}
  \includegraphics[width=1.0\textwidth]{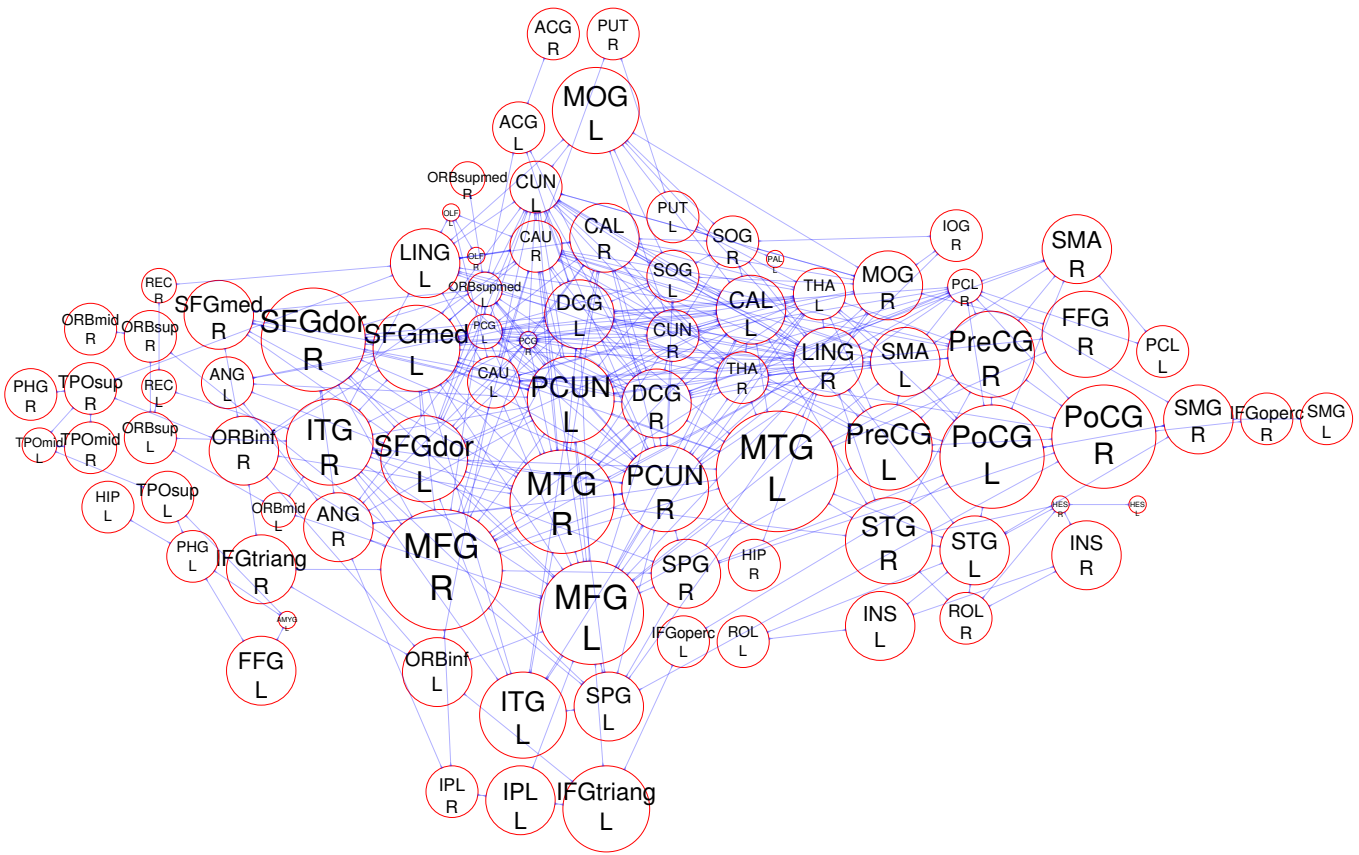}
  \caption[]{\label{fig:brain_layout}
    (Colour online)
    Functional human brain example network 
    (spring model layout of the non-isolated nodes, labels see \cite{Achard2006}).
    Disk area is roughly proportional to node weight (ROI volume).
  }
\end{figure*}

\begin{figure*}
	\begin{center}%
  		\includegraphics[width=\textwidth]{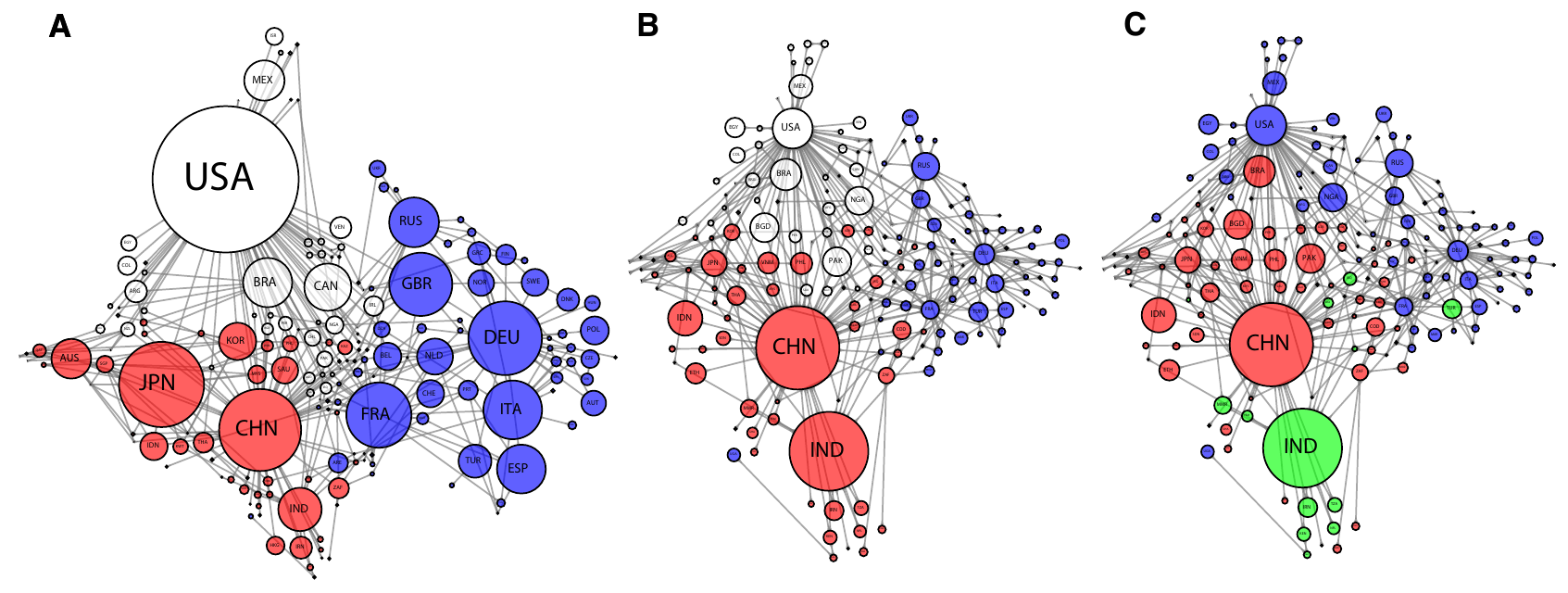}	
	\end{center}
  \caption[]{\label{fig:trade}
    (Colour online)
    World trade network of significant trade relations in 2009
    (spring model layouts, country codes according to ISO 3166).
    Disk area is proportional to 2008 GDP (A) or population (B,C).
    Node colour indicates three-group solutions of Newman's \cite{Newman2006} modularity-based partition algorithm
    (see Appendix B, Modularity).
    The node-weighted (\nsi) version using GDP (A) and 
    the unweighted version (B) give almost identical groups,
    whereas the \nsi\ version using population (C) differs considerably, 
    producing more equally populated regions.
  }
\end{figure*}

Often, a typically finite set of nodes of the studied network is either
explicitly or implicitly considered representative of a much larger finite or
infinite set of objects of interest (which we will call the {\em domain of
interest} or in sampling contexts the {\em population}), either by being a
somehow selected or sampled subset of this larger set or, more often, by
constituting some kind of discretization, aggregation, or coarse-graining of it.
Typical examples are networks of
\begin{enumerate}
  \item functional connections between differently sized {\em regions of interest} (ROIs) in the human brain, 
  		as in \cite{Achard2006,Zhou2006,Hagmann2008} and Fig.\,\ref{fig:brain_layout},
  \item dynamical couplings or statistical associations between time-series measured at discrete regular grid points, 
  in irregular mesh cells,
  or at otherwise sampled {\em discrete locations} on some manifold
  (e.\,g., a climate network either using a latitude-longitude-regular grid on the Earth's surface, 
  as in \cite{Tsonis2004,Donges2009a},
  or with meteorological stations as nodes),
  \item routing connections between {\em autonomous systems} (AS's) in the internet, 
  		representing groups of individual servers, users or ranges of IP addresses, as in \cite{Pastor-Satorras2001,Vazquez2002,Chen2002,Siganos2003},
  \item cross-references between {\em articles} containing different amounts of content in an online encyclopedia, as in \cite{Capocci2006,Zlatic2009},
  \item social relationships between {\em households} consisting of different numbers of individuals, as in \cite{Comola2009,Comola2010},
  \item trade relationships between {\em countries} with differing gross domestic product 
  		and representing different numbers of consumers, as in \cite{Serrano2003} and Fig.\,\ref{fig:trade},
  \item proximities between {\em sampled state vectors} in the (reconstructed) phase space of a dynamical system \cite{Marwan2009,Donner2010,Donner2010b}, 
  		sampled at irregular points in time.
\end{enumerate}
Depending on the meshing, sampling, or parcellation method, the chosen level of aggregation or description, and the availability of data,
some parts of the domain of interest might be represented by relatively more nodes than others,
e.\,g.:
the polar or densely populated regions on the Earth's surface in a climate network 
(since grid points cluster at the poles and meteorological stations cluster in
populated areas); the subcortical area in the human brain (when this is parcellated into smaller regions);
the younger AS's in the internet (usually having below-average numbers of users);
the more technical subjects in the encyclopedia (usually being organized into
shorter articles); the childless population of a village (having a higher ratio of households per people);
in the trade network the industrialized world when the interest is in consumers (consisting of more countries per population)
or the non-industrialized world when the interest is in GDP (consisting of more countries per GDP);
the more densely sampled time periods in the dynamical system.
Often, this problem of over-representing some parts of the domain of interest
is directly related to the size distribution of the objects chosen as nodes.
That distribution is frequently heavy-tailed, e.\,g., for AS's in the internet, articles in the encyclopedia,
and countries in the trade network.

\begin{figure}
  \begin{center}
      \includegraphics[width=\columnwidth]{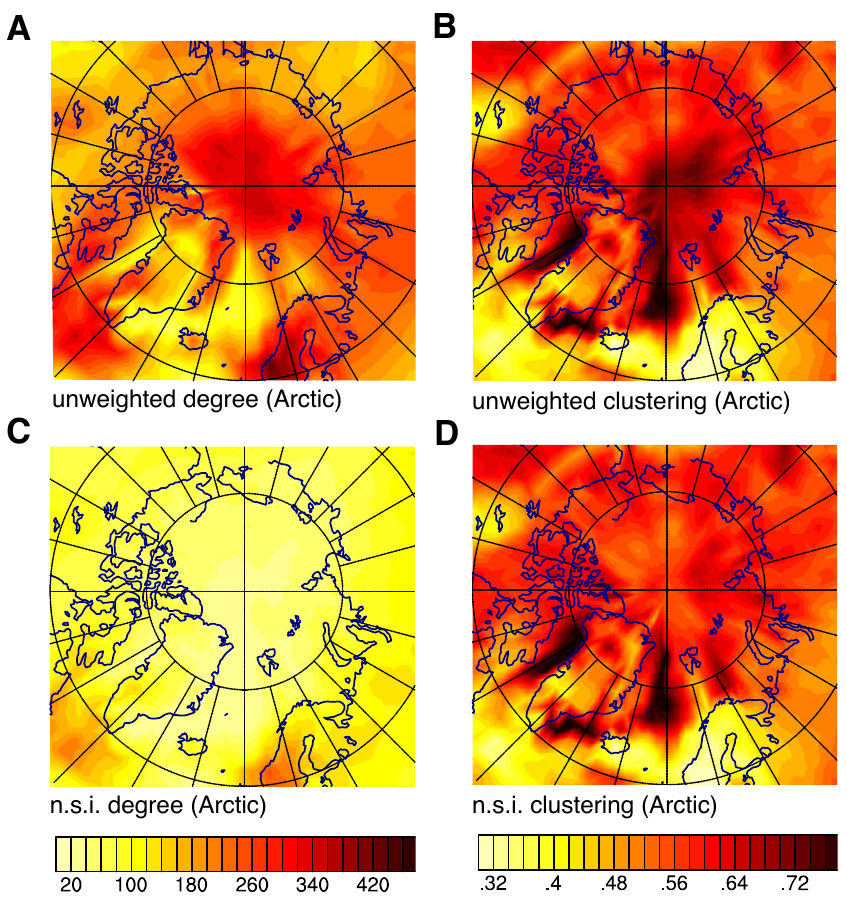}
  \end{center}
  \caption[]{\label{fig:north}
    (Colour online) Comparison of unweighted and weighted (\nsi) versions of
    degree (A,C) and clustering coefficient (B,D) in the northern polar
    region (Lambert equal area projection) of a global climate network
    representing correlations in temperature dynamics. The high values at the
    pole in (A,B) turn out to be an artefact of the increasing grid density
    toward the pole, as demonstrated by (C,D). }
\end{figure}

The described representation bias can cause serious pitfalls in the interpretation of results obtained from the selected network,
if one wants to make inferences about structural and topological properties of the domain of interest
(i.\,e., in the above examples, about
all locations on the globe or in the brain,
all users of the internet,
all units of content in the encyclopedia,
all persons in the village,
all consumers,
or all time points on the trajectory, respectively).
From both a statistical and an approximation-theoretical point of view it is therefore important to
first decide which structural or topological properties of the domain of
interest we are interested in (e.\,g., the connectivity distribution or amount
of clustering), and then to determine what measure in the selected network could be used as a good estimate or approximation of these properties of the underlying domain of interest.
Often, the network construction (that is, the choice of nodes and links)
also involves some parameters like sampling density, grid origin, orientation and size, mesh size, or link inclusion thresholds,
and there are often systematic influences of these parameters on the results of any measurements in the resulting network.
This can lead to selective bias,
as in the case of the dorsal cingulate gyrus in the brain network,
whose betweenness value (a popular measure of node importance) 
depends very much on whether it is treated as one node DCG or as two nodes DCG.L and DCG.R (see Sec.\,\ref{subsec:centrality}).
The effect can even lead to completely artificial features
like in the climate network, as depicted in the centre of Fig.\,\ref{fig:north}\,(A,\,B).

Because of the above observations, our purpose in this paper is to improve the
estimation or approximation power of common network measures by introducing into
their calculation a suitable kind of weighting of all individual nodes. For
this, we use {\em node aggregation weights} based on, e.\,g., ROI volume,
inverse grid density, mesh cell size, inverse sampling density, IP address
ranges of AS's, article length, household size, or a country's population or
GDP. A suitable choice of weights is sometimes difficult (e.\,g., for AS's or
countries) and the weights may have to be estimated (e.\,g., in case of sampled
state vectors). Our focus in this paper, however, is not on the derivation of
suitable weights, but on how to make proper use of them once they are given.

To avoid confusion, we emphasize that there exists a theory of ``weighted
networks'' \cite{Barrat2004} in which {\em links} instead of nodes have weights
representing quantities like length or capacity. But that kind of weights and
the related theory is of little help here since the type of situation we are
concerned with calls for {\em node} weights instead, and we will see that the
corresponding measures differ from those in the other theory, even if the node
weights are somehow translated into link weights. Actually, some real-world
networks (like the world trade network) are probably best described as having
{\em both} node and link weights, as well as being directed.
In this paper, we are treating the case of undirected networks with node weights,
but the same methodology can easily be applied to transform 
network measures that already make use of link weights or link direction
into node-weighted versions that use node weights as well.

\subsection{Climate networks}

To exemplify our approach, assume that the domain of interest is the set of all
points on the Earth's surface and the significant linear correlations between
the surface air temperature time series at pairs of such points. This can be
interpreted as a ``network'' (mathematically, an infinite simple graph) with
uncountably many nodes and (unknown) links. However, we only observe data for a
finite subset of points, say those $64,082$ regular grid points that have
integer latitude and longitude degrees (which is a fairly common grid type in
Earth sciences). Then the set of significant linear correlations between the
temperature time series at these sampled points defines a (finite) {\em climate
network}
\cite{Tsonis2004,Tsonis2006,Tsonis2008,Yamasaki2008,Gozolchiani2008,Donner2008,Donges2009a,Donges2009b}
whose properties, as measured by common network statistics, are somehow hoped to
be representative of related properties of the temperature dynamics on the whole
globe. E.\,g., the degree of a node in the finite network corresponds to how
much surface area this point is correlated to in the whole domain of interest.
When degree is computed in the standard way, however, the resulting large
regional differences will mostly reflect the strongly differing amounts of
surface area each node represents (small area per node at the poles, large area
per node on the equator), instead of indicating ``real'' regional differences in
the connectivity of the underlying domain of interest (the temperature field).

The significance of observed features can often be assessed by comparing results
with those obtainable in a similar ``benchmark'' network in which the links have
been replaced by a spatially homogeneous link distribution, which was done,
e.\,g., in \cite{Henderson2011} to show how the underlying geometry of a
cortical network influences network statistics. Similarly, for our climate
network, a perfectly homogeneous link distribution in the domain of interest
(like the one in which two points are linked iff their angular distance is less
than five degrees) would lead to regional differences of the degree distribution
in the grid-based network, just because its node density and therefore its link
density increase towards the poles (Fig.\,\ref{fig:bylat}, dashed lines). For
the same reason, the standard local clustering coefficient shows an artificial
increase towards the poles. Fig.\,\ref{fig:north}\,(A,\,B) shows such artifacts
in a real-world climate network similar to that of \cite{Donges2009a}. We will
see below that this effect can be avoided by using a node weight proportional to
the inverse node density (= cosine of latitude), and by using the sum of all
neighbour's weights as a measure of degree (this measure being called {\em
area-weighted connectivity} in \cite{Tsonis2006}), and the weighted proportion
of interlinked pairs of neighbours of a node as a measure of clustering, instead
of the classical degree measure and clustering coefficient.

Figure~\ref{fig:north}\,(C,\,D) shows the meaningful regional differences in the
example real-world climate network that remain after the artifacts have been
removed in this way. In that example, we can check the validity of this by
comparing these results to those obtained from a climate network based not on a
latitude-longitude grid but on a ``geodesic'' grid that has an approximately
homogeneous node density all over the globe (see also \cite{Donges2011}, results
not shown here for brevity). Using the classical degree and clustering measures
in the latter, homogeneously sampled network gives results almost identical to
those in Fig.\,\ref{fig:north}\,(C,\,D) rather than (A,\,B). Such a change of
grid, however, usually requires some kind of interpolation of the available
data, which can introduce other problems and is obviously only possible for
specific kinds of network constructions.

\subsection{Outline}

Surprisingly, to our best knowledge, the by now vast literature on complex
networks contains almost no node-weighted measures, except for the
above-mentioned area-weighted connectivity measure, whereas other techniques
(e.\,g., finite elements) and methods of data analysis (e.\,g., empirical
orthogonal functions, a weighted version of principal components analysis
\cite{North1982}) often use some weighting or other adjustment to avoid similar
biases or artifacts.

In this paper, we therefore present a fairly {\em general strategy} for deriving
node-weighted versions of network measures that can be expected to give
estimates or approximations of properties of the domain of interest that are in
a certain sense consistent whenever the network is of the type in which the
links can be interpreted as indicating some kind of similarity or ``close''
relationship, as it is more or less the case in all the cited examples. The
approach will not be directly useful for other kinds of networks, e.\,g., if
links represent some kind of ``complementarity'' instead of similarity, like in
most bipartite networks. Also, it will not be applicable when only the network
as a whole can be considered ``representative'' of the domain of interest, but
when individual nodes cannot be considered representative of some well-defined
part of it, as will often be the case when network sampling methods such as
random node or link sampling or snowball sampling are used (related estimation
problems are treated in \cite{Lee2006,Kolaczykbook2010}).

We then apply this strategy to many of the commonly used network measures and
illustrate the effects in a number of example networks from the above list.
Fortunately, there is a quite simple pragmatic approach to finding useful
weighted versions of network measures which allows us to postpone a more
detailed analysis of the statistical estimation or numerical approximation
properties for further research. This approach is {\em axiomatic} rather than
analytic in that it requires our measures to fulfil an easily verified condition
of {\em node splitting (or twin merging) invariance.}

After stating preliminary matter in Sec.\,\ref{sec:prelim} and giving more
details on our illustrative example networks (Sec.~\ref{sec:examples}), we will introduce
this concept formally and shortly relate it to a statistical and approximation
interpretation in Sec.\,\ref{sec:approaches}. We then proceed with presenting a
comprehensive set of according network measures in Secs.\,\ref{sec:local}--\ref{sec:global}, 
illustrating each one's effect in those example networks 
for which the respective network measure has been considered important in the literature,
but not aiming at analysing each example network with the full set of measures.
We end with a more detailed description of an application to climate networks in
Sec.\,\ref{sec:appl} and a conclusion (Sec.~\ref{sec:conclusion}). In two
Appendices (online), we present some additional measures, give versions of the
new measures that allow for a simpler comparison with their unweighted
counterparts, and shortly discuss how the related parameter of {\em typical
weight} can be estimated.

\section{\label{sec:prelim}%
  Preliminaries
}

\noindent
Let $G=(\nodes,\edges)$ denote a finite undirected simple graph (the {\em network} under consideration) with known
{\em node} or {\em vertex set} $\nodes$,
{\em edge} or {\em link set} $\edges\subseteq\{\{i,j\}:i\neq j\in\nodes\}$,
and {\em adjacency matrix} ${\sf A}=(a_{ij})_{i,j\in \nodes}$, where $a_{ij}\in\{0,1\}$, and $a_{ij}=1$ iff $\{i,j\}\in\edges$.
For simplicity, we assume that $\nodes=\{1,\dots,N\}$ for some natural number $N>1$.
The {\em neighbours} of a node $v\in \nodes$, i.\,e., the members of $v$'s {\em (punctured) neighbourhood}
\begin{align}
  \nodes_v=\{i\in \nodes: a_{iv}=1\}=\{i\in \nodes: a_{vi}=1\}
\end{align}
are those nodes that $v$ is directly {\em linked} to by an edge.
We will also use the {\em extended adjacency matrix} ${\sf A}^+=(a^+_{ij})_{i,j\in \nodes}={\sf A}+{\sf I}$ with
\begin{align}
  a^+_{ij}=a_{ij}+\delta_{ij},
\end{align}
where ${\sf I}=(\delta_{ij})_{ij}$ is the identity matrix and $\delta$ the Kronecker symbol, 
with $\delta_{ij}=1$ for $i=j$ and $\delta_{ij}=0$ for $i\neq j$.
Moreover, we will need the {\em extended} or {\em unpunctured neighbourhood}\footnote{Note that in classical mathematical topology, the term ``neighbourhood of a point'' implies that the point itself is included,
otherwise one speaks of a ``punctured'' neighbourhood.
In the network literature, the term ``neighbour'' also sometimes refers to nodes not directly linked.
In our terminology, a node is not a neighbour of itself but still a member of its unpunctured neighbourhood.}
\begin{align}
  \nodes^+_v = \{i\in\nodes:a^+_{iv}=1\} = \nodes_v\cup\{v\}.
\end{align}

In addition, we will assume that each node $v$ is assigned a positive real-valued {\em (aggregation) weight} $w_v$.
As many of our measures will involve unweighted or weighted means over nodes or pairs of nodes,
we also introduce the {\em total weight}
\begin{align}
  \textstyle W=\sum_{i\in\nodes}w_i
\end{align}
and a shorthand notation for (weighted) averages of functions of nodes or node pairs:
\begin{align}
  \langle g(v)\rangle_v &= \textstyle\frac{1}{N}\sum_{v\in\nodes} g(v),\nonumber\\
  \langle g(v)\rangle_v^w &= \textstyle\frac{1}{W}\sum_{v\in\nodes} w_v g(v),\nonumber\\
  \langle h(i,j)\rangle_{ij} &= \textstyle\frac{1}{N^2}\sum_{i\in\nodes}\sum_{j\in\nodes} h(i,j),\quad\mbox{and}\nonumber\\
  \langle h(i,j)\rangle_{ij}^w &= \textstyle\frac{1}{W^2}\sum_{i\in\nodes}\sum_{j\in\nodes} w_ih(i,j)w_j.
\end{align}

Most of the remaining notation will follow the reviews of 
Newman \cite{Newman2003}, Boccaletti \cite{Boccaletti2006}, and da F.\,Costa \cite{DaFCosta2007}.
In cases where a measure is commonly normalized using a factor of $1/(N-1)$, $1/(N-1)(N-2)$, etc.,
we will use instead the factors $1/N$, $1/N^2$ etc., to keep things simpler.

\begin{figure*}
  \begin{center}
      \includegraphics[width=\textwidth]{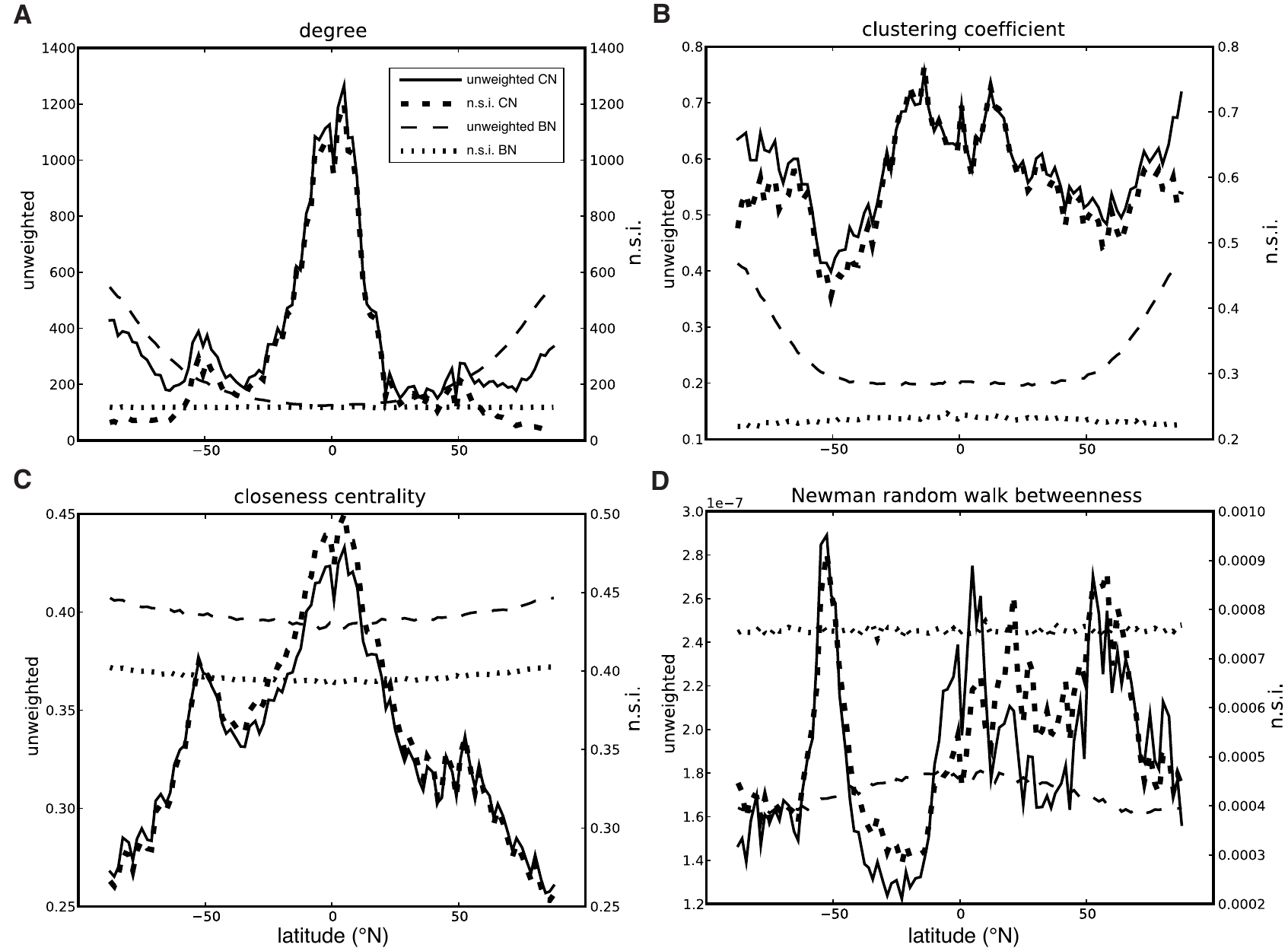}
  \end{center}
  \caption[]{\label{fig:bylat}
    Comparison of unweighted and \nsi\ versions of four common local network measures
    (see Sec.\,\ref{sec:local} for definitions) in a real-world climate network (CN) and a benchmark network (BN).
    (A) degree $k_v,k^\ast_v$, (B) clustering coefficient $C_v,C^\ast_v$,
    (C) closeness centrality $CC_v,CC^\ast_v$, and (D) Newman's random walk betweenness $NB_v,NB^\ast_v$.
    Measures are averaged along bands of equal latitude and plotted against latitude.
    Real-world global climate network representing correlations in surface air temperature dynamics.
    Benchmark network defined on the same grid with independent link probabilities depending on distance alone.
    The benchmark lines show that the observed increase in the unweighted degree
    and clustering coefficient near the poles at $\pm 90^\circ$ latitude is
    mainly due to the vanishing node weight of $\cos$(latitude), whereas the
    effect on closeness centrality is much smaller. In case of Newman's random
    walk betweenness, the slight increase of the unweighted version towards the
    equator in the benchmark network reflects the fact that those nodes
    represent larger surface areas, hence a random walk on the globe will cross
    this area more often. This explains in part why in the unweighted version in
    the real-world network the central peaks are more prominent than in the
    \nsi\ version. }
\end{figure*}

\section{\label{sec:examples}%
    Examples of networks with nodes of different size
}

As stated in previous sections, node weights are ubiquitous for graph
representations of complex systems. We will show the applications of our
weighted measures to various example networks, ranging from the human brain, the
internet, over Wikipedia to world trade. Firstly we describe the details of the
constructed weighted networks. The focus application to climate networks will be
presented individually in Sec.\,\ref{sec:appl}.

\subsection{Human brain}

Functional Magnetic Resonance Imaging (fMRI) time series have been widely used
to study neural activities in the brain from a network perspective. A node is
represented by a cortical region of interest, while a link is often
characterized by some statistical association measuring the correlation between
different regions (i.\,e., linear Pearson correlation, nonlinear mutual
information, or frequency dependent correlation by Wavelets~\cite{Achard2006},
etc.). We consider two regions are connected if their correlation exceeds a
threshold, which can be either based on the correlation values or in terms of
probability under some appropriate null hypothesis. The resulting functional
connectivity of nervous systems has been shown to display high clustering and
short path length which confers a capability for both specialized or modular
processing in local neighbourhoods and distributed or integrated processing over
the entire network~\cite{Achard2006,Hagmann2008}. The cerebral  cortex is a thin
folded sheet tightly confined by the skull and is thus an archetypal example of
a complex network that is strongly constrained by geometry. The understanding of
the properties captured by a variety of network measures (i.\,e., high
clustering, short path length, motifs, and modularity, etc.) has been pointed
out to be very limited because the role of the spatial geometry has been largely
underestimated~\cite{Henderson2011}. Many anatomical features show
distance-dependent properties, e.\,g., the density of corticocortical neural
connections, volume, processing steps, signal travel times, and genetic encoding
needed to specify connectivity.

We re-examine a version of the network of functional connections between
differently sized Region of Interests (ROIs) in the human brain as it was
described in \cite{Achard2006}. After some appropriate preprocessing on the
acquired fMRI data, the network consists of 90 cortical and subcortical time
series extracted from each individual. The resulting graph is shown in
Fig.~\ref{fig:brain_layout}, where the weight of a node is represented by its
associated volume of the ROIs.

\subsection{Internet}

A well-known type of internet mapping is obtained by considering so-called {\em
autonomous systems (AS's)}. On the AS level, each node represents an AS while
each link between two nodes represents the existence of a peer connection among
the corresponding AS's in Border Gateway Protocol (BGP) routing tables.
Traditionally, much interest is in identifying a possible power law
$P(k_v>x)\sim x^{1-\gamma}$ for the degree distribution
\cite{Pastor-Satorras2001,Vazquez2002,Siganos2003}. Power laws have been
reported by different studies of AS maps, and their exponents seem to be stable
over a number of years \cite{Pastor-Satorras2001,Siganos2003}, which could help
to devise a novel class of dynamical models of the internet. However, most
studies use BGP data collected by the Oregon route views project only, which may
provide an incomplete picture of the internet connectivity~\cite{Chen2002}.

Whereas in the literature, usually each AS is treated the same despite the
considerable differences in size (three orders of magnitude), we associate here
with each AS a node weight proportional to the size of the IP address space
allocated to that AS in terms of Classless Inter-Domain Routing (CIDR) prefixes,
a common measure of network size that can be used as an approximation of the
fraction of the internet represented by that AS (although better measures might
be possible). In other words, we consider the set of all IP addresses allocated
by CIDR as our domain of interest $G_0$ which we study be means of a network $G$
of AS's that each represent a certain part of $G_0$. To construct our network,
we used the January 2010 BGP routing table snapshot
(\texttt{http://archive.routeviews.org/oix-route-views/\-2010.01/oix-full-snapshot-2010-01-27-1200.bz2})
from the Oregon route views project, and a corresponding CIDR prefix allocation
snapshot from (\texttt{http://www.cidr-report.org/as2.0/aggr.html}), giving a
network of over 30,000 nodes.

\subsection{Wikipedia} 

Wikipedia is an intriguing research object from a sociologist's point of view:
nodes are articles which are published by a number of independent individuals in
various languages, edges are reference hyperlinks which cover topics they
consider relevant. Several models for the growth of Wikipedia have been proposed
to mimic the plausible preferential attachment mechanisms that might explain the
apparent scale-freeness of the resulting networks~\cite{Capocci2006,Zlatic2009}.

As another example network, we used as nodes all 33,359 articles containing the
word ``physics'' from the 30 July 2010 snapshot of the English language version
of Wikipedia, and made an undirected link between two articles if either
references the other, resulting in an average degree of 31.9. 
Other authors study directed Wikipedia networks \cite{Capocci2006,Zlatic2009}, 
but treating it as an undirected network can be partially justified by the fact
that in Wikipedia one can follow references backwards using the function ``What
links here''. Since the individual articles represent quite different amounts of
text (between one and 283kB), it is straightforward to use their size in characters
as the node weights $w_v$.

\subsection{World trade}

Finally, we also consider for illustration a network of countries where two
countries are linked when they trade considerable amounts (similar to
\cite{Serrano2003}), e.\,g., if the total value of their mutual reported imports
and exports in 2009 accounted for at least $10\%$ of the total reported foreign
trade value of at least one of the two countries, based on data from
comtrade.un.org. Such a network is shown in Fig.\,\ref{fig:trade}. The
topological characterization of the world trade web (WTW) is of primary interest
for the modeling of crisis propagation at the global level, and it has been
reported in \cite{Serrano2003} that the unweighted WTW displays some typical
properties of complex networks, i.\,e., scale free degree distribution,
small-world properties, and high clustering coefficients. As node weights, we
use either population or gross domestic product in 2008 (as reported by the
IMF), both showing considerable differences.

A much more realistic model of the world trade network would of course 
use weighted and directed links representing actual imports and exports, 
in addition to node weights, so one cannot attach much real-world importance to
the exemplary results that we will present here for illustration with this
simplified network.

\section{\label{sec:approaches}%
  Approaches to node weighting in network measurements
}

\subsection{\label{subsec:statistical}%
  Statistical interpretation
}

\noindent
In many cases the nodes $\nodes$ and links $\edges$ of the studied network $G$ are simply a subset from 
a larger (maybe infinite) network $G_0$ with nodes $\nodes_0$ and unknown links $\edges_0$
that constitute the domain of interest whose structural and topological properties we are interested in.
Since often $\nodes_0$ is a manifold like the Earth's surface, the brain volume, or a phase space,
we will call the elements of $\nodes_0$ {\em points} here 
although the following consideration also applies to discrete sets $\nodes_0$ like that of all households in a society.
If we can interpret the node set $\nodes$ to be a {\em sample} from the {\em population} of points $\nodes_0$,
resulting from some sampling procedure,
then we might adopt a classical statistical approach and consider any measurements in the sample network $G$ 
as estimates of certain statistics of the population network $G_0$ that we are truly interested in.
E.\,g., the classical measure of {\em degree} of a node $v\in\nodes$,
\begin{align}
  k_v=k_v(G)=|\nodes_v| = \textstyle\sum_{i\in\nodes} a_{iv},
\end{align}
could be interpreted as the simplest estimator of 
the number (or proportion) $k_0(v)$ of points in $\nodes_0$ to which $v$ is linked in $G_0$.
If, however, the sampling procedure is such that certain points $v\in\nodes_0$ are selected for the sample 
with a higher individual {\em sampling probability} $p_v$ than others, 
basic statistics tells us that a much better estimator of $k_0(v)$ 
is a weighted sum,
\begin{align}
  \tilde k_v = \textstyle\sum_{i\in \nodes_v} w_i = \sum_{i\in\nodes} w_i a_{iv}
\end{align}
with suitable node weights $w_v\geqslant 0$.
As in the well-known Horvitz-Thompson estimator of a sample mean,
the optimal weights $w_v$ are given by inverse probability weighting, 
i.\,e., they are inversely proportional to the sampling probabilities, $w_v\propto 1/p_v$.
E.\,g., if $G$ is a climate network constructed from meteorological stations 
and the probability $p_v$ of having a station in location $v$ is proportional to the local human population density,
then any analysis of $G$ should assign a node at $v$ a weight proportional to the inverse human population density,
to make sure that the climates in sparsely and densely populated areas are equally represented.
In some cases, statistical considerations can also motivate more sophisticated 
choices of the weights $w_v$, like the reliability-adjusted Kriging weights 
used for meteorological station data in \cite[Eq.\,15]{Rohde2011}.

If we were to follow this statistical approach more thoroughly,
we would try to identify each property of the domain of interest we are interested in
with some statistics $f_0$ of $G_0$, and then use a suitably weighted estimator $\tilde f$ of $f_0$
that is at least statistically {\em consistent} (i.\,e., converges to $f_0$ in a certain sense as $N$ increases),
hopefully also {\em unbiased} (i.\,e., has an average error of zero) and {\em efficient} (i.\,e., has small variance),
and maybe even {\em robust} (i.\,e., is not very sensitive to only local changes in the network).
Verifying these properties for a large number of different network measures
is however a research program requiring much analytical effort 
beyond the scope of a single paper. 
Moreover, it would likely require some complicated continuity assumptions on $G_0$
that would restrict the applicability of those measures considerably,
which is why we pursue a simpler approach here to find suitable weighting schemes 
for individual network measures.

Note that in principle, the above estimator $\tilde k_v$ 
can be interpreted as a special case of the {\em strength} $s_v=\sum_{i\in\nodes_v} w_{vi}$
of a node $v$ in a {\em directed, (link-)weighted network} in which we simply use 
the node weight $w_i$ of the target node as the link weight $w_{vi}$ of the directed link.
For other network measures, this interpretation is, however, 
not helpful since the measure might have no counterpart for directed weighted networks, 
or that counterpart is unsuitable for our problem (as in case of the clustering coefficient, see below).

\subsection{\label{subsec:numerical}%
  Numerical approximation
}

\noindent
If the domain of interest $G_0$ provides some notion of (geometric) distance 
(like any spatially embedded network does),  
an alternative approach is to consider 
each node $v\in\nodes$ as representative for a small cell
$\cell_v$ of points in $v$'s geometrical vicinity in $\nodes_0$,
whose size (in terms of some suitable measure, e.\,g., Lebesgue measure) we denote by $w_v$.
By {\em geometrical vicinity} we mean those points of the underlying domain of interest
that have a small geometrical distance from $v$,
as opposed to its {\em neighbourhood} $\nodes^+_v$ that consists of those nodes in the network with a network-theoretic distance $\leqslant 1$.
This interpretation would be adequate, e.\,g., 
if $\nodes_0$ is a continuous manifold and $\nodes$ a subset of points
on a grid or derived by some meshing procedure (e.\,g., adaptive mesh refinement, \cite{Plewa2005}).

If, because of the continuity properties of the underlying system, 
it can be expected that all nodes $v'\in \cell_v$ are linked to more or less the same nodes in $G_0$ as $v$ is,
then a natural approximation for an interesting statistics $f_0$ of $G_0$ would use the aggregation weight $w_v$
wherever the formula for $f_0$ involves the node $v$.

\begin{figure}
  \begin{center}
    \includegraphics[width=0.9\columnwidth]{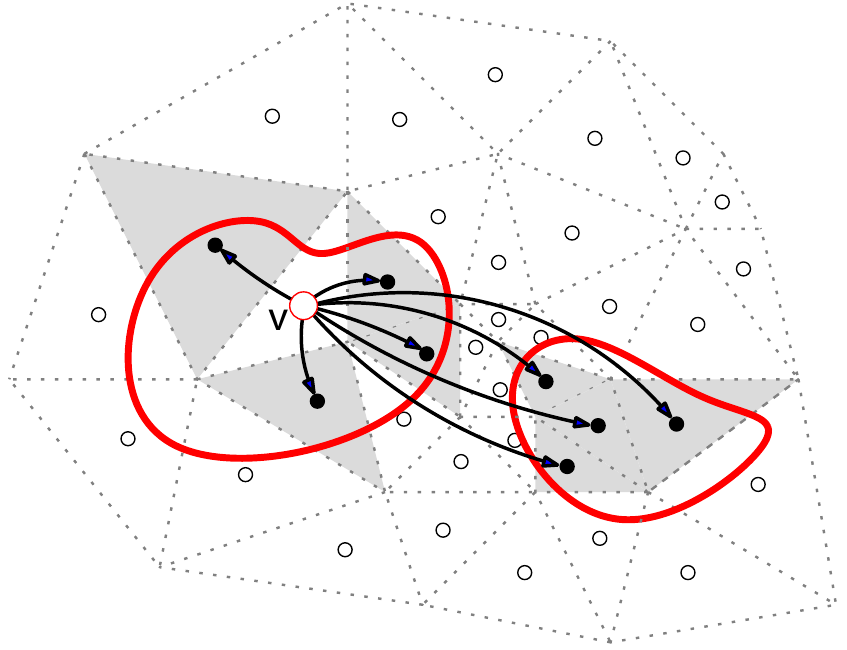}
  \end{center}
  \caption{\label{fig:approx}
    (Colour online)
    A set $\nodes$ of nodes (circles) representing cells of different size of the domain of interest (dashed triangles).
    In many applications, a node $v$ will be linked to one or more smoothly bounded regions of the domain of interest (red surrounded regions),
    containing $v$ itself and some other nodes (filled circles; arrows show the resulting links in the network $G$).
    The red surrounded area of size $k_0(v)$ can then be approximated by the grey shaded area of size $\tilde k_v$,
    or more accurately by the grey shaded area plus $v$'s own cell area, giving the estimate $k^\ast_v$.
    The classical degree $k_v$ is just the number of filled circles.
  }
\end{figure}

E.\,g., for the above measure of degree,
we could again use $\tilde k_v = \sum_{i\in \nodes_v}w_i$ instead of $k_v=|\nodes_v|$ to approximate $k_0(v)$,
since each node $i$ in $v$'s $G$-neighbourhood represents $w_i$ ``many'' nodes in $G_0$
of which most can be expected to be linked to $v$ as well
(Fig.\,\ref{fig:approx} illustrates this idea).
In the context of climate networks,
each node $i$ represents a portion of the Earth's surface of relative size $w_i = \cos($latitude of $i)$,
and $\tilde k_v$ is known as {\em area weighted connectivity} \cite{Tsonis2006}.

In many cases of continuous domains of interest $G_0$,
a point $v\in \nodes_0$ is usually also linked to all or at least most of the points in its geometrical vicinity.
More formally, in many cases the following {\em local connectedness} condition will hold for some suitable distance function $d$:
for all $v\in \nodes_0$ there is $\varepsilon>0$ such that each point $i\in \nodes_0\setminus\{v\}$ with $d(i,v)<\varepsilon$ is linked to $v$.
E.\,g., if $\nodes_0$ is the Earth's surface,
and $i,j\in \nodes_0$ are linked in $G_0$ iff the surface temperature time series for $i$ and $j$
exhibit a product-moment correlation coefficient larger than some threshold value,
then the smoothness of the underlying physics will imply the above.

In such a network, an alternative approximation to $k_0(v)$ would then be
\begin{align}
  k^\ast_v = \tilde k_v + w_v = \textstyle\sum_{i\in \nodes^+_v}w_i = \textstyle\sum_{i\in\nodes} w_i a^+_{iv},
\end{align}
which can be expected to be a better approximation if the mesh is fine enough.
This estimator can also be interpreted as a classical numerical approximation to
the integral of the indicator function of $v$'s unpunctured neighbourhood in $G_0$
(see Fig.\,\ref{fig:approx} again).

However, for many of the more complex network measures we will study below, e.\,g., random walk based measures or spectra,
it will not be possible that easily to interpret our weighted versions of those measures as approximations to integrals,
and a thorough analysis of their approximation qualities would require much technical effort.
This is why we rather pursue a third, more pragmatic approach,
which is motivated by a simple property that both the statistical and the numerical approximation interpretations have in common.

\subsection{\label{subsec:pragmatic}%
  Pragmatic axiomatic approach
}

\noindent
In both the statistical and the approximation approach,
it is clear that the estimation or approximation should usually become better
when the resolution of $G$ as a description of the domain of interest $G_0$ is increased by
replacing some or all nodes by a larger set of nodes representing smaller parts of $G_0$.
Such {\em refinements} would usually change the corresponding inverse sampling densities
or cell sizes that we use as our aggregation weights $w_v$.
Let us now consider the case of sufficiently high resolution, i.\,e., 
where the sample is dense enough or the cell sizes are small enough to resolve 
all structural features of $G_0$ that are considered relevant,
so that we do not expect there to be considerable inhomogeneities inside the region of $G_0$ 
represented by each individual node.
Now imagine that an {\em elementary refinement} of $\nodes$ was performed in which
only one old node $s\in\nodes$ was replaced by two new nodes $s',s''\in \nodes_0$
which together represent more or less the same subset of $\nodes_0$ as $s$ did.
Then this would leave the aggregation weights $w_i$ of the other nodes $i\in\nodes\setminus\{s,s',s''\}$ mostly unchanged,
whereas the weights $w_{s'}$ and $w_{s''}$ of the new nodes would approximately sum up to the former $w_s$.
Also, because the resolution was assumed to be sufficient already, 
$s'$ and $s''$ would be linked to more or less the same nodes as $s$ was, and most likely also to each other.
In that case, 
a good estimate or approximation $\tilde f$ to some statistic $f_0$ of $G_0$ should probably become somewhat more precise,
but should certainly not be changed much by such an elementary refinement.
This intuitive reasoning can be turned into a simple axiomatic guiding requirement
when we idealize the above situation as follows.

\begin{figure}
  \begin{center}
    \includegraphics[width=0.9\columnwidth]{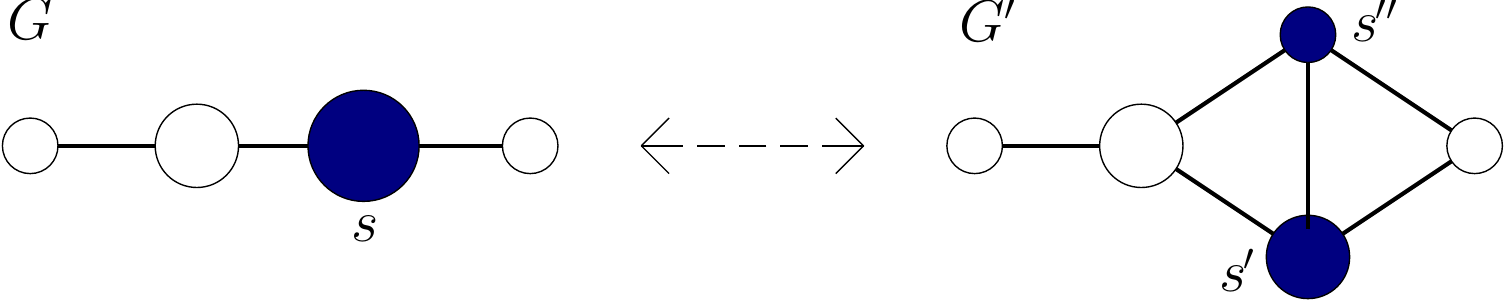}
  \end{center}
  \caption{\label{fig:splitting}
    The operation of {\em node splitting} replaces a node $s$ of weight $w_s$
    in network $G$
    with two linked nodes $s',s''$ of weights $w_{s'}+w_{s''}=w_s$
    which get the same neighbourhood as $s$ had, giving network $G'$.
    The inverse operation of {\em twin merging} transforms $G'$ back into $G$.
  }
\end{figure}

Let $G=(\nodes,\edges)$ be a simple graph with weights $w_i>0$ for all $i\in\nodes$,
and let $s\in\nodes$ be some ``old'' node, $s',s''\notin\nodes$ two ``new'' nodes,
$w_{s'},w_{s''}>0$ their weights, and $w_{s'}+w_{s''}=w_s$.
Then the ``refined" graph $G'=(\nodes',\edges')$ with
\begin{align}
  \nodes' &= \nodes\setminus\{s\}\cup\{s',s''\} \nonumber\\
  \mbox{and}\quad
  \edges' &= \big\{\{i,j\}\in \edges:i,j\neq s\big\}\\
      &\quad {}\cup\big\{\{i,s'\},\{i,s''\}:\{i,s\}\in \edges\big\} \cup \big\{\{s',s''\}\big\}\nonumber
\end{align}
and with weights $w_i$ for all $i\in\nodes'$ will be called a {\em node splitting refinement} of $G$.
That is, $G'$ is derived from $G$ by ``splitting'' the node $s$ into two new interlinked nodes $s',s''$ with the same total weight as $s$
and linking those two to exactly those nodes to which $s$ was linked (see Fig.\ \ref{fig:splitting}).

Now a (global) network measure $f$ will be called {\em node splitting invariant (\nsi)}
iff
\begin{align}
  f(G')=f(G)
\end{align}
for all pairs of networks $G,G'$ and all weight functions $w$ for which $G'$ is a node splitting refinement of $G$.
Likewise, if $f_i$ is a network measure which is defined for nodes $i$,
then $f_i$ will be called {\em node splitting invariant} iff
\begin{align}
  f_i(G')=f_i(G)\quad\mbox{and}\quad f_{s'}(G')=f_{s''}(G')=f_s(G)
\end{align}
for all such $G,G',w$ and all $i\in\nodes\setminus\{s,s',s''\}$.
Similar definitions are possible for measures with more than one argument but are not needed here.
In other words, a \nsi\ measure is {\em unaffected by any node splitting refinements.}
This is, however, not to say that we actually perform any refinements or node splittings when applying a node splitting invariant measure.
Rather, the notion of node splitting must be thought of as only a hypothetical, idealized operation that could be applied in principle,
and the consistency requirement of node splitting invariance is only to make sure that the node weights are used in a proper way
that is suitable for links which represent some kind of ``similarity''.
In other words, to {\em define} a \nsi\ measure only requires us to analyse what {\em would} happen if nodes were split in the suggested way.
To {\em apply} a \nsi\ measure to a given network, 
one can just use the corresponding formula as one would with an unweighted measure, 
without any need to think about node splitting. 

In the context of coarse-graining, 
node splitting invariance can also be called {\em twin merging invariance},
requiring that $f$ should not change 
whenever two twins $s',s''$ with weights $w_{s'},w_{s''}$ are merged into 
one node $s$ with weight $w_s=w_{s'}+w_{s''}$,
where two linked nodes $s',s''$ are called {\em twins} iff they have the same neighbourhood,
$\nodes^+_{s'}=\nodes^+_{s''}$.
While node splitting corresponds to an idealized form of refinement,
twin merging is an idealized form of coarse-graining and is the inverse operation of node splitting, 
hence both notions of invariance are equivalent.

While the above definitions of node splitting and twin merging 
are suitable for networks in which links represent a kind of similarity or direct connection,
for which it is natural to require that twins are linked and to assume that the parts of a split node are linked,
other types of networks might call for a different definition of node splitting and twin merging.
If, e.\,g., links represent some kind of ``complementarity'' instead of similarity,
a natural kind of splitting would leave $s'$ and $s''$ unlinked,
and the corresponding definition of ``twin'' would require that twins are not linked.
Such a variant would lead to weighted network measures similar but not identical to ours,
but we do not pursue this in the present paper.

~

An example of a \nsi~measure is the above-defined {\em \nsi~degree}
$k^\ast_v = \sum_{i\in \nodes^+_v}w_i$,
and that it is indeed \nsi\ is easily seen from the fact that
\begin{align}
  \nodes^+_{s'}(G')=\nodes^+_{s''}(G') &= \nodes^+_s(G)\setminus\{s\}\cup\{s',s''\}, \nonumber\\
  \nodes^+_i(G') &= \nodes^+_i(G)\setminus\{s\}\cup\{s',s''\},
  \\
  \mbox{and}\quad
  \nodes^+_j(G') &= \nodes^+_j(G) \nonumber
\end{align}
for all $i\in \nodes_s(G)$ and $j\in\nodes\setminus \nodes^+_s(G)$.

Note that the definition of node splitting invariance aka twin merging invariance 
does not at all rely on the formal specification of an underlying domain of interest $G_0$,
but it depends on the network $G$ and the weights $w_i$ alone,
which makes this tool much easier to use than estimation theory or approximation theory.
Nevertheless, a conjecture and working hypothesis of this paper is that
\nsi\ measures are the natural candidates for good estimation or approximation
of the corresponding properties of a potentially underlying domain of interest,
and that they will usually prove to be statistically consistent and exhibit good convergence properties
when the domain of interest and the sampling or meshing procedures fulfil some suitable continuity or measurability properties
and when the aggregation weights $w_i$ are chosen accordingly.

In the following, we will therefore present \nsi\ versions $f^\ast$
of a number of local and global network measures $f$ that can be found in the literature,
and we will refer to the possibly underlying domain of interest only when motivating some interpretations of these measures,
but without formally defining that statistics $f_0$ of $G_0$ which is supposed to be estimated or approximated by $f^\ast$.

The basic construction mechanisms we will use are
\begin{enumerate}
  \item[(i)]    to sum up aggregation weights wherever the original measure counts nodes,
  \item[(ii)]   to use unpunctured neighbourhoods $\nodes^+_v$
                wherever the original measure uses punctured neighbourhoods $\nodes_v$
                (in other words, to consider $v$ as linked to itself),
  \item[(iii)]  to also allow for equality of $i,j$ wherever the original measure involves a sum over distinct nodes $i,j$, and
  \item[(iv)]   to ``plug-in'' a \nsi~version of a measure $g$ wherever this $g$ is used in the definition of another measure $f$.
\end{enumerate}
Both mechanisms (i) and (ii) were used in the definition of $k^\ast_v$ above,
and an example for mechanisms (iii) and (iv) will be given in the following section
when we will consider the clustering coefficient.

\section{\label{sec:local}%
  Local measures
}

\noindent
A network measure $f_v=f_v(G)$ that is defined for each node $v\in\nodes$ will be called {\em local} here.
(Note that we always understand the terms ``neighbour'' and ``local'' as referring to the network topology,
not to some possibly underlying geometry. So, local neighbours might be geometrically far apart.)

\subsection{Degree}

\noindent
We already treated the {\em degree} measure $k_v=|\nodes_v|$
and defined the {\em \nsi\ degree} of $v$ as
\begin{align}
  k^\ast_v = \textstyle\sum_{i\in \nodes^+_v}w_i.
\end{align}

Let us compare $k_v$ and $k^\ast_v$ in two example networks.

In our human brain example network, 
the nodes with the highest degree $k_v$ are the right lingual gyrus and the left precuneus region 
(LING.R and PCUN.L, see \cite{Achard2006} for these abbreviations), 
connected to 32 and 31 of the other 89 nodes, 
i.\,e., to about one third of the nodes.
The volume of the 90 individual ROIs varies by a factor of 23 (Fig.\,\ref{fig:brain_layout})  
If it is used as a node weight for \nsi~degree, 
the right lingual gyrus and the left precuneus again have the largest values, 
but these are now $k^\ast_v\approx 0.46 W$, showing that in reality, 
both regions are functionally connected not to one third but rather to almost half the entire brain (in terms of volume).
The third most connected node in terms of $k_v^\ast$ (volume) is the left middle frontal gyrus 
(which seems consistent with other measures of node importance as reported in \cite{Achard2006}),
but in terms of $k_v$ (nodes) it is the left calcarine cortex (CAL.L).
Since the number of linked nodes basically depends on the level of detail in the used parcellation of the brain,
$k_v$ can change considerably for different parcellations even if the ROI represented by node $v$ remains unchanged, 
and $k_v$ seems to be influenced much more than $k^\ast_v$ by the choice of parcellation.  

\begin{figure}
  \includegraphics[width=\columnwidth]{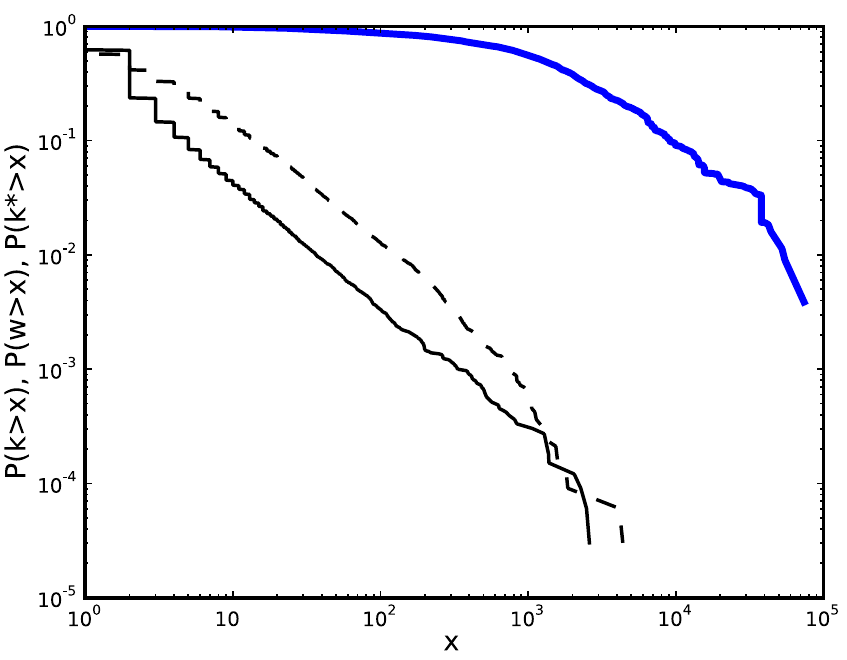}
  \caption[]{\label{fig:as_kloglog}
  	(Colour online)
    Log-log plot of the complementary cumulative distribution function of 
    degree $k_v$ (thin black line),
    node weight $w_v$ (dashed line),
    and \nsi~degree $k^\ast_v$ (thick blue line)
    in the routing network of autonomous systems in the internet.
    Power laws would appear linear.
  }
\end{figure}

In the internet example network, 
the earlier findings of power laws for the degree distribution 
reported in \cite{Pastor-Satorras2001,Vazquez2002,Siganos2003}
are supported 
by the apparent linear relationship in the log-log plot in Fig.\,\ref{fig:as_kloglog} (thin black line).
In other words, the distribution of the number $k_v$ of linked AS's of a given AS $v$ seems to follow a power law.
However, also the distribution of the size $w_v$ of a given AS $v$ in terms of CIDR prefixes seems to follow a power law,
as can also be seen in Fig.\,\ref{fig:as_kloglog} (dashed line),
and the linear correlation coefficient between $\ln k_v$ and $\ln w_v$ is high ($\approx 0.5$),
so the power law for $k_v$ might be a consequence of the one for $w_v$.
If we ask for the share of the internet (instead of the number of AS's) a given AS $v$ is linked to,
$k^\ast_v/W$ seems a more accurate estimate of this than $k_v/N$, and the findings are different:
The AS with highest $k_v$ also has the highest $k^\ast_v$, 
but while it is linked to only $7.9\%$ of all AS's (since $k_v=0.079 N$), 
it seems to be linked to approx.~one fourth of all IP adresses (since $k^\ast_v=0.24 W$).
When plotting the probability that a randomly chosen CIDR prefix belongs to an AS 
that is linked to other AS's with more than $x$ total CIDR prefixes,
this does no longer show a clear power law behaviour (Fig.\,\ref{fig:as_kloglog}, thick blue line).
Also, $\ln k^\ast_v$ is less strongly correlated ($\approx 0.3$) to $\ln w_v$ than $\ln k_v$ is,
and less strongly correlated ($\approx 0.4$) to $\ln k_v$ than $\ln w_v$ is. 

\subsection{Clustering coefficient}

\noindent
The {\em local clustering coefficient} of $v$,
\begin{align}
  C_v
  = \frac{\sum_{i\in \nodes_v}\sum_{j\in \nodes_v} a_{ij}}{k_v(k_v-1)}
  = \frac{N^2\langle a_{vi}a_{ij}a_{jv}\rangle_{ij}}{k_v(k_v-1)}
  \in [0,1],
\end{align}
is the probability that two nodes drawn at random from those linked to $v$ are linked with each other.
We get a weighted version by employing all four mechanisms (i)--(iv),
giving the {\em \nsi\ local clustering coefficient}
\begin{align}
  C^\ast_v
    = \frac{W^2\langle a^+_{vi}a^+_{ij}a^+_{jv}\rangle_{ij}^w}{k^{\ast 2}_v}
    \in \left[\frac{w_v(2 k^\ast_v - w_v)}{k^{\ast 2}_v},1\right]
    \subseteq[0,1],
\end{align}
which estimates the probability that two weight units (or {\em points} in terms of $G_0$) 
drawn at random from the part of the network linked to $v$ are linked with each other.
Because we use ${\sf A}^+$, $C^\ast_v$ tends to be larger than $C_v$ if $k_v$ is small, 
and it is defined for all nodes while $C_v$ only makes sense when $k_v>1$.
If the weights vary considerably, 
$C^\ast_v$ and $C_v$ can rank the nodes quite differently,
although neither needs to be significantly linearly correlated with $w_v$.
$C^\ast_v$ and $C_v$ can also differ considerably when the weights {\em inside} 
$\nodes_v$ vary strongly.

\begin{figure}
  \includegraphics[width=\columnwidth]{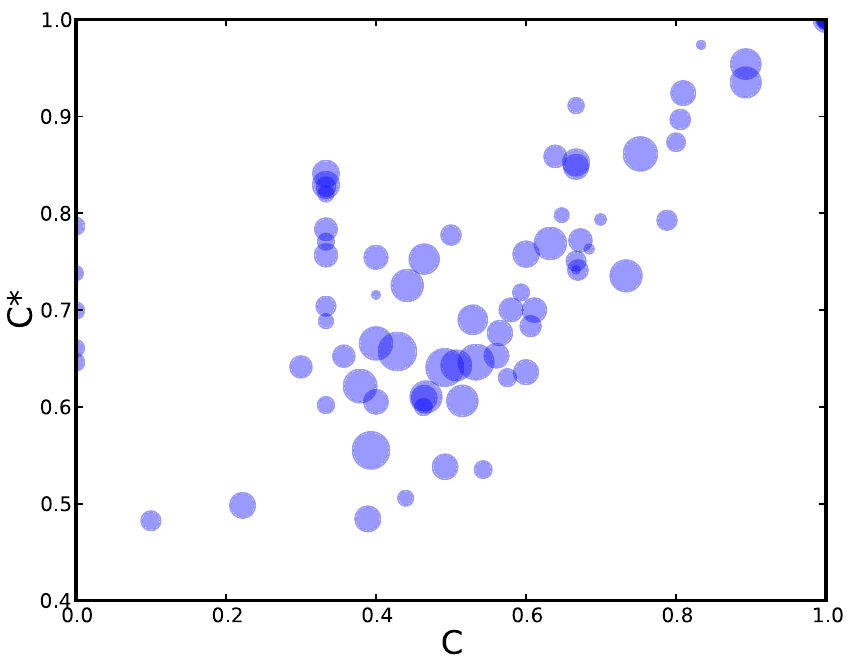}
  \caption[]{\label{fig:brain_C}
    Clustering coefficient $C_v$ vs.~\nsi\ clustering coefficient $C^\ast_v$
    of those nodes in the functional human brain network that have at least two neighbours,
    showing considerable differences (Pearson correlation is $\varrho=0.65$).
    Disk area is proportional to node weight (ROI volume). 
  }
\end{figure}

All these effects can be seen nicely in our brain example network (Fig.\,\ref{fig:brain_C}).
The right thalamus region (THA.R), e.\,g., is in the top half (rank 37) according to its $C_v$ of $0.54$, 
but almost at the bottom (rank 86 out of 90) according to its $C^\ast_v$ of $0.53$, 
although the absolute values are almost equal.
This is because among its 21 neighbours, 
it is rather the smaller ones (like THA.L and CAU.R) that are linked with many other neighbours.

One might think that similar to the case of degree, 
also a directed version of the formula for the local clustering coefficient from the theory of (link-)weighted networks
\cite{Barrat2004},
which is $c^w_v= \sum_{i\in \nodes_v}\sum_{j\in \nodes_v} a_{ij}(w_{vi}+w_{vj})/2s_v(k_v-1)$,
could be a good candidate if the directed link-weights are defined as $w_{vi}=w_i$,
giving $c^w_v= \sum_{i\in \nodes_v}\sum_{j\in \nodes_v} a_{ij}(w_i+w_j)/2s_v(k_v-1)$.
But it is easy to see that this does not behave well under node splitting or twin merging
since a linked pair $i$---$j$ contributes $w_i+w_j$ instead of $w_iw_j$.
If, e.\,g., $\nodes_v=\{i,j,s',s''\}$ with $w_i=w_j=w_{s'}=w_{s''}=1$, $a_{ij}=0$,
and $s',s''$ are twins linked to $i$ but not to $j$, then $c^w_v=C_v=3/6$,
but after merging the twins $s',s''$ into one new node $s$ with $w_s=2$, 
one has $k_v=3$, $s_v=4$, and $c^w_v=(w_i+w_s)/2s_v(k_v-1)=3/16$,
much smaller than before. 
In contrast, $C^\ast_v=19/25$ before and after the twin merging.

\subsection{\label{subsec:centrality}%
  Measures of centrality and betweenness
}

\noindent
Many measures try to assess several aspects of ``node importance''.
Based on the distances of one node $v$ to all others, we consider three variants
of {\em closeness centrality,}
\begin{align}
  CC_v = 1/\langle d_{vi}\rangle_i,\quad
  CC'_v = \langle 2^{-d_{vi}}\rangle_i,\quad
  CC''_v = \langle 1/d_{vi}\rangle_i,
\end{align}
the latter also being called the {\em efficiency} of $v$,
where $d_{vi}$ is the number of links on a shortest path from $v$ to $i$,
or, if there is no such path, either $\infty$ or $N$, depending on the convention chosen
\cite{DaFCosta2007,Dangalchev2006,Stephenson1989,Latora2002}.
A weighted version of $CC_v$ should give us the inverse average distance of $v$ 
from other weight units or points rather than from other nodes.
But for this to become \nsi,
one has to interpret (somewhat peculiarly) each node to have unit (instead of zero) distance to itself.
This is because after an imagined split $s\to s',s''$, the two parts $s',s''$ of $s$ have unit not zero distance.
The {\em \nsi\ distance function} is hence given by
\begin{align}
  d^\ast_{vv} = 1 \quad\mbox{and}\quad d^\ast_{vi}=d_{vi}\mbox{~for~}i\neq v,
\end{align}
i.\,e., the zeros on the diagonal of the ordinary distance matrix are replaced by ones to get the \nsi\ distance matrix,
without changing any off-diagonal entries. 
Using this, we can derive $v$'s {\em \nsi\ closeness centrality} measures as
\begin{align}
  CC_v^\ast = \frac 1 {\langle d^\ast_{vi}\rangle_i^w} 
  			= \frac{W}{\sum_{i\in\nodes} w_i d^\ast_{vi}} 
  			= \frac{W}{w_v+\sum_{i\in\nodes} w_i d_{vi}},
\end{align}
$CC'^\ast_v = \langle 2^{-d^\ast_{vi}}\rangle_i^w$,
and $CC''^\ast_v = \langle 1/d^\ast_{vi}\rangle_i^w$.
All take values in $[0,1]$.

\begin{figure}
  \includegraphics[width=\columnwidth]{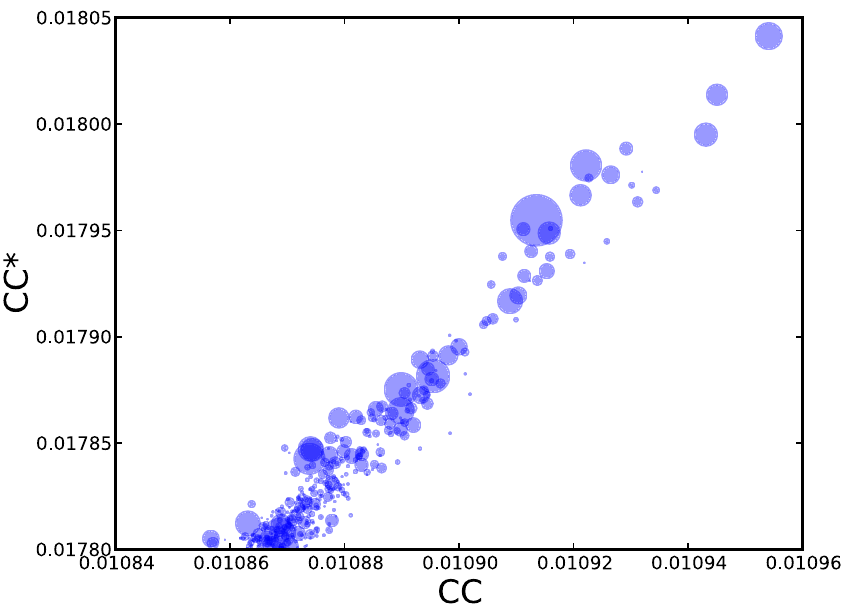}
  \caption[]{\label{fig:as_CC}
    Closeness $CC_v$ vs.~\nsi\ closeness $CC^\ast_v$
    of those nodes in the internet network (see text) with the highest values.
    Disk area is proportional to node weight (no.~of CIDR prefixes). 
  }
\end{figure}

In the internet example network, $CC_v$ and $CC^\ast_v$ are generally quite small,
vary only little, and are highly correlated.
Still, both measures lead to different rankings of the most central nodes (Fig.\,\ref{fig:as_CC}).

\begin{figure}
  \includegraphics[width=\columnwidth]{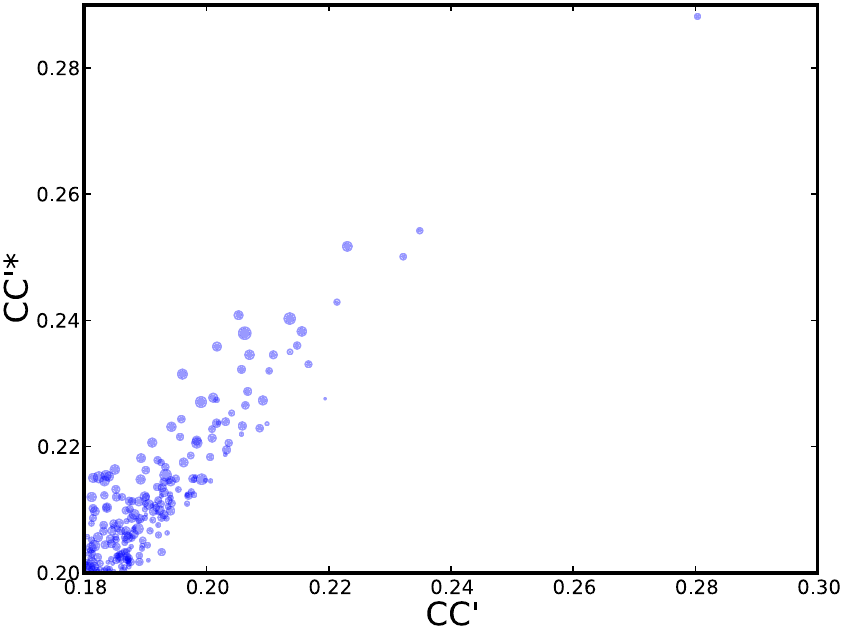}
  \caption[]{\label{fig:wiki_eCC}
    Exponential closeness $CC'_v$ vs.~its \nsi\ version $CC'^\ast_v$
    of those Wikipedia articles on physics (see text) with the highest values.
    Disk area is proportional to node weight (article size in characters).
  }
\end{figure}

Also in our Wikipedia example network, 
$CC_v$ and $CC^\ast_v$ were not very discriminatory, 
but $CC'_v$ and $CC'^\ast_v$ were.
Fig.\,\ref{fig:wiki_eCC} depicts their relationship for the most central nodes,
again showing considerable differences in ranks.
According to $CC'_v$, the top ten central articles in decreasing order are those named
``Physics'', ``Mathematics'', ``Chemistry'', ``Germany'', ``Science'', 
``Physicist'', ``Quantum mechanics'', ``Albert Einstein'', ``Astronomy'', and ``Engineering'',
while according to $CC'^\ast_v$ the list is
``Physics'', ``Mathematics'', ``Germany'', ``Chemistry'', ``Science'', 
``Japan'', ``Italy'', ``Albert Einstein'', ``Russia'', and ``Astronomy''.
``Physicist'' for instance has the sixth highest value of $CC'_v$ but only the 21th largest value of $CC'^\ast_v$
because it is linked to a very large number ($7.6\%$ of the nodes) 
of comparatively short articles (accounting for $5.5\%$ of the total text) on individual scientists,
so that $CC'_v$ treats them as a larger part of the network than $CC'^\ast_v$ does .

~

\begin{figure}
  \begin{center}
    \includegraphics[width=0.5\columnwidth]{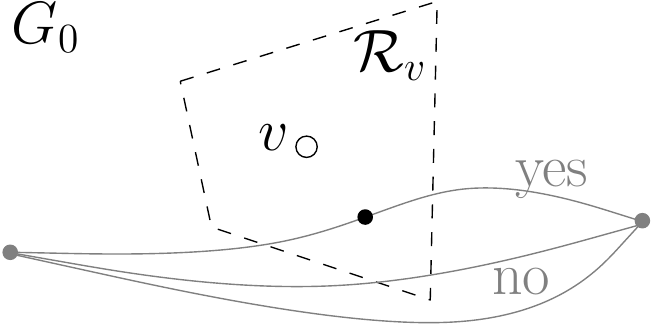}
  \end{center}
  \caption{\label{fig:bc}
    Interpretation of \nsi\ shortest path betweenness as the probability density
    that a randomly chosen shortest path (grey curves) between two randomly chosen points (grey dots)
    in the domain of interest $G_0$
    passes through a randomly chosen point (black dot)
    in the area $\cell_v$ (dashed region) represented by $v$ (black circle).
  }
\end{figure}

\noindent
Other, somewhat more sophisticated importance measures depend on the paths between 
all other nodes that lead through a given node $v$.
The {\em (shortest path) betweenness} of $v$ is the proportion of shortest paths between randomly chosen nodes $a,b$
that lead through $v$:
\begin{align}
  BC_v = \langle n_{ab}(v)/n_{ab}\rangle_{ab}  \in [0,1],
\end{align}
where $n_{ab}$ is the total number of shortest paths from $a$ to $b$,
and $n_{ab}(v)$ is the number of those paths that pass through $v$ as an inner node.
Formally, $n_{ab}$ can be written as a sum over all node tuples $(t_0,\dots,t_{d_{ab}})$ with $t_0=a$ and $t_{d_{ab}}=b$,
where the summands are either zero or one, depending on whether each $t_\ell$ is linked to its successor $t_{\ell+1}$,
for $\ell=0,\dots, d_{ab}-1$.
As the latter condition can be written as a product of elements of the adjacency matrix, we have:
\begin{align}
  n_{ab} = \sum_{\substack{(t_0,\dots,t_{d_{ab}})\in \nodes^{d_{ab}+1}\\ t_0=a,~t_{d_{ab}}=b}}
    \textstyle\prod_{\ell=1}^{d_{ab}}a_{t_{\ell-1} t_{\ell}}.
\end{align}
A similar formula holds for $n_{ab}(v)$, only that for some $m$ in $1\dots d_{ab}-1$, $t_m$ must equal $v$:
\begin{align}
  n_{ab}(v) =   {\textstyle\sum_{m=1}^{d_{ab}-1}}
    \sum_{\substack{(t_0,\dots,t_{d_{ab}})\in \nodes^{d_{ab}+1}\\ t_0=a,~t_m=v,~t_{d_{ab}}=b}}
		\textstyle\prod_{\ell=1}^{d_{ab}}a_{t_{\ell-1} t_{\ell}}.
\end{align}
When $s$ is hypothetically split into $s'+s''$, any shortest path through $s$ becomes a pair of shortest paths, one through $s'$ and the other through $s''$.
Also, a shortest path from $s''$ to some $b\neq s'$ will never meet $s'$.
Thus, to make $BC_v$ \nsi, it suffices to make $n_{ab}$ and $n_{ab}(v)$ \nsi\ by making each path's contribution
proportional to the weight of each inner node (that is, to the product of these weights!),
where in case of $n_{ab}(v)$ we have to skip $w_v$ in this product:
\begin{align}
  n^\ast_{ab} &= \sum_{\text{(as above)}}
    a_{t_0 t_1}\textstyle\prod_{\ell=2}^{d_{ab}}(w_{t_{\ell-1}}a_{t_{\ell-1} t_{\ell}}),\nonumber\\
  n^\ast_{ab}(v) &= {\textstyle\frac 1{w_v}}{\textstyle\sum_{m=1}^{d_{ab}-1}}
    \sum_{\text{(as above)}}
    \Big(a_{t_0 t_1}\textstyle\prod_{\ell=2}^{d_{ab}}(w_{t_{\ell-1}}a_{t_{\ell-1} t_{\ell}})\Big).
\end{align}
The {\em \nsi\ shortest path betweenness}
\begin{align}
  BC^\ast_v = \langle n^\ast_{ab}(v)/n^\ast_{ab}\rangle_{ab}^w \in [0,1/w_v]
\end{align}
can then be interpreted as an estimate of the probability (or probability density)
that a randomly chosen shortest path between two randomly chosen points in the underlying domain of interest $G_0$
passes through {\em a specific randomly chosen point in the area $\cell_v$} represented by $v$, as illustrated in Fig.\,\ref{fig:bc}.
The product $w_vBC^\ast_v$ then estimates the probability that such a path passes through {\em any} point in $\cell_v$,
which is not \nsi\ but is additive under node splitting:
$w_{s'}BC^\ast_{s'}+w_{s''}BC^\ast_{s''}=w_{s}BC^\ast_{s}$.

Newman \cite{Newman2001} gives an $O(|\nodes||\edges|)$-time algorithm to compute $BC_v$ for all $v$,
based on Dijkstra's algorithm,
and this can easily be adapted to compute $BC^\ast_v$ for all $v$
with the same algorithmic time complexity.

In our human brain example network, according to $BC_v$ the top ten ROIs in decreasing order are the nodes 
TPOsup.R, LING.L, LING.R, TPOmid.L, MFG.R, TPOsup.L, MFG.L, STG.L, CAU.R, and ORBinf.R,
while according to $BC^\ast_v$ the list is
TPOmid.L, TPOsup.R, LING.R, LING.L, TPOsup.L, PCL.R, PHG.L, THA.R, CAU.R, and STG.L.
The left and right middle frontal gyrus (MFG) are missing in the latter list (having ranks 13 and 19)
because most shortest paths that lead through them are between relatively small nodes, 
(left and bottom region in Fig.\,\ref{fig:brain_layout}), 
whereas the right thalamus (THA.R) has a high degree but many of its larger neighbours are not linked to each other, 
leading to a large value of $BC^\ast_v$.
In this network one can also nicely see the effect of network design choices 
on network statistics.
If we slightly modify the parcellation and treat the mid-sized left and right parts of the 
dorsal cingulate gyrus as one node $s$ instead of two 
($s'=$DCG.L and $s''=$DCG.R, having quite similar neighbourhoods),
leaving the rest of the network unchanged,
then their $BC$-values increase considerably 
from $BC_{s'}=0.029$ and $BC_{s''}=0.036$ to $BC_s=0.053$, or from 24th and 18th rank to 11th rank.
Their $BC^\ast$-values, however, behave rather nicely in that the new
value $BC^\ast_s=.0000144$ (rank 19) is approximately the average of the two old values
$BC^\ast_{s'}=0.0000114$ (rank 23) and $BC^\ast_{s''}=0.0000171$ (rank 17).
Had DCG.L/R been exact twins, $BC^\ast$ would not have changed at all.

More centrality and betweenness measures will be discussed in the next subsection and in Sec.\,\ref{subsec:spectral}.

\subsection{\label{subsec:randomwalks}%
  Measures based on random walks
}

\noindent
Several network measures are based upon the idea of a random walk along the links of a network without isolated nodes,
with all neighbours $j$ of a node $i$ having the same transition probability
\begin{align}
  p_{ij} = a_{ij}/k_i.
\end{align}
The crucial ideas in constructing \nsi\ versions of such measures are
to make $p_{ij}$ proportional to $w_j$ and to allow the walk to stay at node $i$
(which now also allows for the existence of isolated nodes):
\begin{align}
  p^\ast_{ij} = a^+_{ij}w_j/k^\ast_i.
\end{align}
Such a walk can be thought to approximate a random walk in $G_0$ which moves to each linked point
with the same probability.
If $G_0$ is a continuous domain, this discrete random walk must not be confused with a continuous Wiener process, however.
In particular, its individual steps might bridge long distances in terms of the domain's geometry.

With the above transition probabilities,
the probability of visiting or staying inside $\{s',s''\}$ after a split $s\to s'+s''$ is
$p^\ast_{vs'}+p^\ast_{vs''}=p^\ast_{vs}$ or
$p^\ast_{s's'}+p^\ast_{s's''}=p^\ast_{s''s'}+p^\ast_{s''s''}=p^\ast_{ss}$, respectively,
which means the walk is not influenced by the split.
With the original {\em transition matrix} ${\sf P}=(p_{ij})_{ij}$,
the equilibrium distribution and hence the long-time average relative visiting frequencies are given by
$p_v=k_v/K$, where $K=N\langle k_v\rangle_v$ is twice the number of links in $G$.
With the {\em \nsi\ transition 
matrix} ${\sf P}^\ast=(p^\ast_{ij})_{ij}$, 
the equilibrium distribution is $p^\ast_v=k^\ast_v/K^\ast$, 
where $K^\ast=W\langle k^\ast_v\rangle_v^w$. 

~

\noindent
The {\em Arenas-type random walk betweenness} of $v$, 
motivated by \cite{Arenas2003} and based on the idea of searching a target,
is the expected number of visits to $v$ on a random walk that starts and ends at some randomly chosen nodes $a$, $b$:
\begin{align}
  AB_v = \langle AB_{av}(b)\rangle_{ab}
\end{align}
with ${\sf AB}(b) = (AB_{ij}(b))_{ij} = \sum_{t=1}^{\infty} {\sf P}'(b)^t$.
Since the walk is assumed to stop as soon as $b$ is reached,
the ``transition'' matrix is here ${\sf P}'(b) = (p'_{ij}(b))_{ij}$ with
\begin{align}
  p'_{ij}(b) = (1-\delta_{bi})a_{ij}/k_i.
\end{align}
The main problem in attaining node splitting invariance is the stopping condition introduced by the term $-\delta_{bi}$.
If the target node $b$ is split into $b'+b''$, and if $b'$ is the new target node,
the walk must not continue after reaching $b''$ since otherwise $AB_v$ would increase.
Hence the walk must sometimes stop earlier than before,
at least when reaching a twin of the target.
As exact twins are usually rare in large networks, 
it seems natural to adopt a somewhat more continuous stopping condition that
may stop the walk with some probability as soon as it enters the neighbourhood of the target.
Using a suitable {\em \nsi\ similarity measure} $\sigma^\ast(i,j)\in[0,1]$
that equals zero for unlinked nodes and one for twins,
we can then define a {\em \nsi\ Arenas-type random walk betweenness}:
\begin{align}
  AB^\ast_v = \langle AB^\ast_{av}(b)\rangle_{ab}^w/w_v
\end{align}
with ${\sf AB}^\ast(b) = (AB^\ast_{ij}(b))_{ij} = \sum_{t=1}^{\infty} {\sf P}'^\ast(b)^t$
and ``transition'' matrix ${\sf P}'^\ast(b) = (p'^\ast_{ij}(b))_{ij}$, where
\begin{align}
  p'^\ast_{ij}(b) = (1-\sigma^\ast(b,i))a^+_{ij}w_j/k^\ast_i.
\end{align}
For the similarity measure we may, e.\,g., use one out of the increasing sequence
\begin{align}
  \sigma^\ast_I(i,j) &= \delta_{\nodes^+_i\nodes^+_j}, \nonumber\\
  \sigma^\ast_{II}(i,j) &= a^+_{ij}w(\nodes^+_i\cap \nodes^+_j)/w(\nodes^+_i\cup \nodes^+_j), \nonumber\\
  \sigma^\ast_{III}(i,j) &= a^+_{ij}w(\nodes^+_i\cap \nodes^+_j)/\max(k^\ast_i,k^\ast_j), \nonumber\\
  \sigma^\ast_{IV}(i,j) &= a^+_{ij}w(\nodes^+_i\cap \nodes^+_j)\cdot 2/(k^\ast_i+k^\ast_j),\\
  \sigma^\ast_{V}(i,j) &= a^+_{ij}w(\nodes^+_i\cap \nodes^+_j)/\sqrt{k^\ast_ik^\ast_j}, \nonumber\\
  \sigma^\ast_{VI}(i,j) &= a^+_{ij}w(\nodes^+_i\cap \nodes^+_j)/\min(k^\ast_i,k^\ast_j), \nonumber\\
  \sigma^\ast_{VII}(i,j) &= a^+_{ij} \nonumber
\end{align}
with $w(M)=\sum_{v\in M}w_v$ for $M\subseteq N$.
In all these versions, the walk may stop with some probability when a more or less twin-like neighbour of $b$ is reached.
If nodes can be expected to ``know'' their neighbours with some probability, 
this stopping behaviour can also be interpreted as meaning that once a neighbour of the target is reached, 
the path to the target will more or less likely be known, 
hence the target can be considered to be found and the random search walk can stop. 

In the brain example network, both the unweighted and \nsi\ version 
(using $\sigma^\ast_{VII}$) of Arenas-type random walk betweenness 
give the same set of nodes with top ten betweenness values, 
but in quite different order: for $AB_v$ it is 
LING.R, PCUN.L, CAL.L, PCUN.R, MFG.L, MTG.R, DCG.L, DCG.R, SFGmed.L, and MFG.R,
while for $AB^\ast_v$ it is
PCUN.L, LING.R, MFG.L, CAL.L, MTG.R, PCUN.R, SFGmed.L, DCG.L, MFG.R, and DCG.R.

In our world trade example network, when we use as node weights the considerably varying 
countries' gross domestic product in 2008 (as reported by the IMF),
then according to $AB_v$ the countries with the largest betweenness are
CHN, USA, DEU, FRA, JPN, IND, RUS, ITA, SGP, and KOR,
whereas according to $AB^\ast_v$ the ranking is much different,
USA, CHN, JPN, KOR, PAN, SAU, MYS, PHL, VNM, and GAB, in descending order.
The last six are all connected to all three of the heavy-weight nodes USA, CHN, and JPN (or FRA in case of GAB),
which explains their high $AB^\ast_v$-values.
Also in the layout in Fig.\,\ref{fig:trade}, they are more centrally located than those in the first list.
Germany (DEU) and France (FRA) on the other hand, 
are missing from the second list mainly because they are connected to neither of those three big economies directly.
A much more realistic model of the world trade network would of course 
use weighted and directed links representing actual imports and exports, 
in addition to node weights,
so one cannot attach much real-world importance to the above exemplary results.

~

\begin{figure}
  \begin{center}
    \includegraphics[width=\columnwidth]{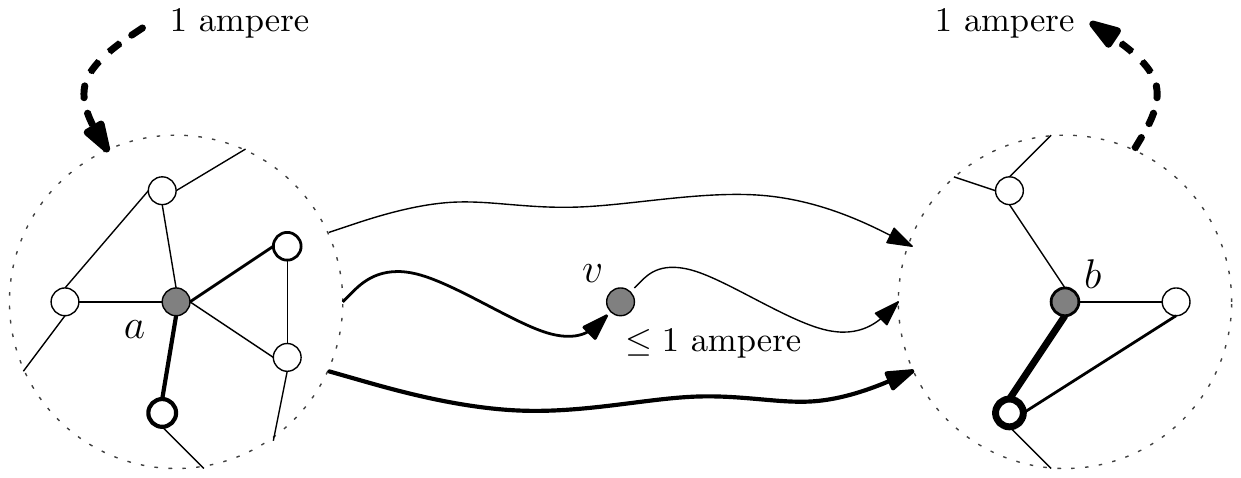}
  \end{center}
  \caption{\label{fig:current}
    Interpretation of \nsi\ Newman-type random walk betweenness as the expected (electric) current flowing through $v$
    when a unit current flows between the neighbourhoods of two randomly selected nodes $a$ and $b$
    and each link's conductance is proportional to both its ends' weights (here represented by line thickness).
  }
\end{figure}

\noindent
In the preceding type of random walk betwenness, each individual visit of the walk to $v$ is counted.
Newman \cite{Newman2005} introduced a similar measure in which, however,
only the ``net'' flow of the walk along each link is considered.
A more intuitive interpretation of his measure is the expected effective current $I_{ab}(v)$ passing through $v$
when the network is interpreted as an electric circuit with all links having unit conductance,
and a unit current is sent from a random node $a$ to a random node $b$.
This explains the definition of {\em Newman's random walk betweenness}
\begin{align}
  NB_v &= \langle I_{ab}(v)\rangle_{ab}\\
  \mbox{with}\quad
  I_{ab}(i) &= \textstyle\frac 1 2\sum_{j\in \nodes_i}|V_j(a,b)-V_i(a,b)|, \nonumber
\end{align}
where the ``electric potential'' vector $\vec V(a,b)$ is given by Kirchhoff's equations
\begin{align}
  {\sf\Lambda}\vec V(a,b) = \vec\delta(a)-\vec\delta(b)\quad\mbox{with}\quad
  \delta_i(c) = \delta_{ic},
\end{align}
where ${\sf\Lambda}={\rm diag}(\vec k)-{\sf A}$ is the {\em Laplacian} matrix of $G$
which will be studied more closely in Sec.\,\ref{subsec:spectral}.
As with the stopping condition above, here the main problem in finding an \nsi\ version
is that the current leaves the circuit at a single node $b$,
so that when $b$ is split into $b'+b''$ and the current leaves at $b'$,
the twin $b''$ will have a different electric potential than $b'$.
This can be overcome in the same way as above,
by using some \nsi\ similarity measure $\sigma^\ast$
and letting a part of the current that is proportional to $w_i\sigma^\ast(b,i)$
leave the circuit also at each neighbour $i$ of $b$.
A similar thing must also be done at the $a$-side where the current enters.
Finally, each link $i$---$j$ must get conductance $w_iw_j$
to reflect the fact that the link $i$---$j$ represents a bundle of links in $G_0$ between
the $w_i$ many points represented by $i$ and the $w_j$ many points represented by $j$.
In order to avoid a dominating influence of the degree on our measure,
we also restrict $a$ and $b$ to nodes not directly linked to $v$,
although this would not be necessary to achieve node splitting invariance.
For the choice $\sigma^\ast=\sigma^\ast_{VII}$ in which the current enters
and leaves at all nodes in the neighbourhood of $a$ and $b$ to some extent,
this approach is depicted in Fig.\,\ref{fig:current}.

All these considerations lead to our definition of {\em \nsi\ Newman-type random walk betweenness}
\begin{align}
  NB^\ast_v &= \langle (1-a^+_{av}){\sf I}^\ast_{ab}(v)(1-a^+_{vb})\rangle_{ab}^w\\
  \mbox{with}\quad
  {\sf I}^\ast_{ab}(i) &= \textstyle\frac 1 2\sum_{j\in \nodes^+_i}w_j|V^\ast_j(a,b)-V^\ast_i(a,b)|, \nonumber
\end{align}
where $\vec V^\ast(a,b)$ is given by the new equations
\begin{align}
  & {\sf D}_w{\sf\Lambda^\ast}\vec V^\ast(a,b) = \vec\delta'(a)-\vec\delta'(b),\nonumber\\
  & \delta'_i(c) = w_i\sigma^\ast(i,c)/\textstyle\sum_{j\in \nodes^+_c}w_j\sigma^\ast(j,c),
\end{align}
and where
\begin{align}
  {\sf D}_w = {\rm diag}(\vec w)\quad\mbox{and}\quad
  {\sf\Lambda^\ast} = {\rm diag}(\vec k^\ast)-{\sf A}^+{\sf D}_w
\end{align}
are the diagonal matrix of aggregation weights and the \nsi\ Laplacian matrix (see Sec.\,\ref{subsec:spectral}).
In this way, after a split $s\to s'+s''$, all potentials $V^\ast_i(a,b)$ with $i\neq s$ remain unchanged,
and $V^\ast_{s'}(a,b)$ and $V^\ast_{s''}(a,b)$ equal the former $V^\ast_s(a,b)$.
The measure $NB^\ast_v$ estimates the expected effective current
passing through each unit-sized part of the region $\cell_v$ of $G_0$ that is represented by $v$,
when the domain of interest $G_0$ is interpreted as an electric circuit with all links having unit conductance,
and a unit current is sent between two random points of $G_0$.

In the brain example network, 
this time the top ten lists according to the unweighted and \nsi\ versions 
of Newman's betweenness measure differ more than for Arenas-type betweenness:
We get 
TPOsup.R, LING.L, LING.R, ORBinf.R, MFG.R, MFG.L, TPOmid.L, CAU.R, STG.L, and ITG.R
according to $NB_v$, but
TPOmid.L, TPOsup.R, TPOsup.L, LING.L, ORBinf.R, TPOmid.R, CAU.R, THA.L, STG.L, and THA.R
according to $NB^\ast_v$.
For example, MFG.L/R is again missing in the second list while THA.L/R is included,
and the explanation is the same as for the case of shortest path betweenness 
(Sec. \ref{subsec:centrality}). 

Newman-type random walk betweenness is also of interest in climate networks, 
see Sec.\,\ref{sec:appl}. 

\section{\label{sec:global}%
  Global measures
}

\subsection{\label{subsec:aggregate}%
  Aggregate measures
}

\noindent
Popular aggregate network statistics include the {\em global clustering coefficient}
$C = \langle C_v\rangle_v \in [0,1]$ with \nsi\ version 
\begin{align}
  C^\ast = \langle C^\ast_v\rangle_v^w \in [0,1],
\end{align}
the {\em global transitivity}
$T = \langle a_{vi}a_{ij}a_{jv}\rangle_{vij}/\langle a_{vi}(1-\delta_{ij})a_{jv}\rangle_{vij}$
which is closely related to $C^\ast$ and has the \nsi\ version 
\begin{align}
  T^\ast = \frac{\langle a^+_{vi}a^+_{ij}a^+_{jv}\rangle_{vij}^w}{\langle a^+_{iv}a^+_{vj}\rangle_{vij}^w} \in [0,1],
\end{align}
the {\em link density} $\varrho = \langle a_{ij}\rangle_{ij} \in [0,1)$,
whose \nsi\ version is 
\begin{align}\textstyle
  \varrho^\ast = \langle a^+_{ij}\rangle_{ij}^w = \frac 1 W\langle k^\ast_v\rangle_v^w \in [0,1]
\end{align}
the {\em average (geodesic) path length} or {\em mean geodesic distance}
$L = \langle d_{ij}\rangle_{ij}\geqslant 0$
with
\begin{align}\textstyle
  L^\ast = \langle d^\ast_{ij}\rangle_{ij}^w>0,
\end{align}
and the {\em global efficiency} \cite{Latora2002} $E = \langle 1/d_{ij}\rangle_{i\neq j} \in [0,1]$
for which we can use
\begin{align}\textstyle
  E^\ast = \langle 1/d^\ast_{ij}\rangle_{ij}^w \in [0,1].
\end{align}
Many authors, e.\,g., compare $C$ and $L$ to their values for randomly rewired graphs to assess the ``small-worldness'' of the network
in a way that is quite sensitive to small structural changes in the network (see \cite{Bialonski2010}),
and using $C^\ast$ and $L^\ast$ for this task should at least reduce that part of this non-robustness 
that is related to node selection and aggregation.

In our internet example network, we get $C=0.21$, i.\,e.,
for a randomly chosen AS $v$,
two randomly chosen other AS's that are both linked to $v$ 
are also linked with each other with probability $21\%$.
If the AS network is constructed more meticulously than here, combining several data sources,
somewhat larger values around $0.4$ are found \cite{Barcelo2004}.
On the other hand, we have $C^\ast=0.8$, meaning that
for a randomly chosen IP address $x$,
two randomly chosen other IP addresses $y,z$ that can be reached from $x$ in at most one routing step,
can also reach each other in at most one routing step with $80\%$ probability.
One reason for the large difference between $C$ and $C^\ast$ is the low average degree 
of only $\langle k_v\rangle_v=4.24$, 
which means that many of the above pairs $y,z$ belong to the same AS and therefore contribute to $C^\ast$ but not to $C$.
Also, we get $E\approx 0.27$ and $E^\ast\approx 0.336$, 
which can be interpreted as indicating that 
a typical distance between two randomly chosen AS's is $1/0.27\approx 3.7$ routing steps,
while between two randomly chosen IP adresses it is only $1/0.336\approx 2.98$, which is $20\%$ less,
showing that large AS's tend to be more central than small ones.

\subsection{\label{subsec:spectral}%
  Characteristic matrices and spectral measures
}

\noindent
Spectral network analysis deals with the eigenvalues and eigenspaces of
characteristic matrices such as the adjacency matrix ${\sf A}$
or the {\em Laplacian} and {\em normal matrices}
\begin{align} {\sf\Lambda}={\rm diag}(\vec k)-{\sf A},\quad {\sf T}={\rm diag}(\vec k)^{-1}{\sf A}, \end{align}
of which we will only consider the first two here.
Spectral analysis can be used to study the centrality of nodes and the community structure of the network,
and to find natural partitions or node classification trees.

~

\noindent
Depending on which matrix properties are considered essential, 
there are at least two different ways in which these matrices can be made \nsi\ in some sense.
One way is to multiply both the rows and columns with the square root of the corresponding node weights,
similar to what is done in a different context 
in the estimation of empirical orthogonal functions from gridded data (e.\,g., \cite{North1982}),
thereby preserving the symmetry of the matrix.
For the adjacency matrix, this gives
\begin{align}
  {\sf A}^{\ast\prime} &= {\sf D}_w^{1/2} {\sf A}^+{\sf D}_w^{1/2},
\end{align}
which has the property that any solution to the eigenequation ${\sf A}^{\ast\prime}\vec x=\lambda\vec x$ is \nsi\
in the sense that after a node splitting $s\to s'+s''$,
the resulting new ${\sf A}^{\ast\prime}$ still has the eigenvalue $\lambda$ with some new eigenvector $\vec y$,
the entries $x_v$ of the eigenvector $\vec x$ that belong to non-split nodes $v\neq s$ remain unchanged,
$y_v=x_v$,
and the quotient of eigenvector entry and square root of the weight is invariant also for the split nodes,
$y_{s'}/\sqrt{w_{s'}} = y_{s''}/\sqrt{w_{s''}} = x_s/\sqrt{w_s}$.
In particular, $\vec y$ has the same $\ell_2$-norm as $\vec x$.
In a sense, the splitting of node $s$ results in 
a corresponding ``splitting'' of all the eigenvectors' $s$-dimension.

After the split, ${\sf A}^{\ast\prime}$ has an $(N+1)$st eigenvalue of zero, 
corresponding to the eigenvector $\vec z$ with
$z_{s'} = \sqrt{w_{s''}}$, $z_{s''} = -\sqrt{w_{s'}}$, and $z_v=0$ for $v\neq s',s''$.
In particular, the {\em $m$-th moment} $\frac 1 N \sum_i\lambda_i^m$ of ${\sf A}^{\ast\prime}$ becomes \nsi\
if the normalization $1/N$ is replaced by the \nsi\ normalization $1/W$: $\frac 1 W \sum_i\lambda_i^m$.
(As $i$ does not refer to a node in this sum, we do not need to weight $\lambda_i^m$ with $w_i$.
The sum itself is already \nsi\ since all non-zero $\lambda_i$ are.)

An alternative and simpler possibility is to use the node weights only for the columns and put
\begin{align}
  {\sf A}^\ast &= {\sf A}^+{\sf D}_w,
\end{align}
destroying symmetry and getting different eigenvectors $\vec y$ as with ${\sf A}^{\ast\prime}$,
although the eigenvalues are the same.
The entries of the eigenvector $\vec x$ that belong to non-split nodes $i\neq s$ still remain unchanged,
but this time $s'$ and $s''$ directly inherit their eigenvector entries from $s$, 
$y_{s'} = y_{s''} = x_s$,
so the new eigenvector $\vec y$ has the same $\ell_\infty$-norm as $\vec x$.

~

\noindent
As both ${\sf A}^\ast$ and ${{\sf A}^\ast}'$ are non-negative,
the Perron-Frobenius theorem guarantees the existence of a non-negative eigenvalue 
of largest absolute value and corresponding non-negative real eigenvectors 
$\vec x^\ast$ and $\vec x^{\ast\prime}$.
Both can therefore be used to get the \nsi\ version of another popular centrality measure:
The {\em (adjacency) eigenvector centrality} $EC_v$ of $v$ is 
the entry $x_v$ of the (non-negative) eigenvector $\vec x$
for ${\sf A}$'s largest eigenvalue,
where $\vec x$ is usually normalized so that $\max_v x_v=1$.
The above shows that we can define {\em \nsi\ eigenvector centrality} as
$EC^\ast_v=x^{\ast}_v=x^{\ast\prime}_v/\sqrt{w_v}$.
A similar measure based on the normal matrix ${\sf T}$ 
is closely related to a major web search engine's page rank measure.

In the brain example network, the top ten central nodes according to $EC_v$ are
PCUN.L, LING.R, CAL.L, PCUN.R, DCG.L, DCG.R, MTG.R, SFGmed.L, CUN.L, and PCG.R,
and according to $EC^\ast_v$ they are
PCUN.L, MTG.R, MFG.L, SFGmed.L, PCUN.R, DCG.L, LING.R, DCG.R, CAL.L, and MFG.R.
Curiously, this time the second rather than the first ranking includes MFG.L/R 
which in this case is probably just because they have 
a much larger weight than CUN.L and PCG.R (which are only in the first ranking).

~

\noindent
For the Laplacian matrix $\sf\Lambda$, also both constructions are possible,
leading to
\begin{align}
  {\sf \Lambda}^\ast = {\rm diag}(\vec k^\ast)-{\sf A}^\ast
  \quad\mbox{and}\quad
  {\sf \Lambda}^{\ast\prime} = {\rm diag}(\vec k^\ast)-{\sf A}^{\ast\prime}.
\end{align}
While the row sums of ${\sf \Lambda}^{\ast}$ are zero just like for $\sf\Lambda$,
those of ${\sf \Lambda}^{\ast\prime}$ need not equal zero,
but the latter matrix is symmetric.
Still, both matrices have the same eigenvalues, 
and those and their eigenvectors behave in the same way under node splitting as those of 
${\sf A}^\ast$ and ${{\sf A}^\ast}'$,
except that the additional $(N+1)$st eigenvalue is now $k^\ast_s$ instead of zero.

The non-symmetric version ${\sf \Lambda}^\ast$ can also be interpreted as a
Laplacian matrix of a directed network in which the links instead of the nodes are weighted,
with link weights $w_{ij}=w_i$.
If the network is the result of the spectral coarse graining procedure described in \cite{Gfeller2008},  
${\sf \Lambda}^\ast$ equals the Laplacian derived in \cite[Eq.~3]{Gfeller2008}. 

~

\noindent
The {\em spectral bisection method} of Fiedler \cite{Fiedler1973} uses
the signs of the eigenvector of the smallest positive eigenvalue of ${\sf \Lambda}$ to find the
most distinguishable two groups of nodes.
Using either ${\sf \Lambda}^\ast$ or ${\sf \Lambda}^{\ast\prime}$
in the same way will provide the {\em \nsi\ spectral bisection}, 
both giving the same result since only the sign of the eigenvector entries matters.

An enhanced version of spectral bisection uses eigenvectors of 
Newman's \cite{Newman2006} generalized modularity matrix to iteratively find communities.
Fig.\,\ref{fig:trade} shows three-group solutions found by the unweighted and \nsi\ versions 
of that algorithm, as described in Appendix B (online only), 
in the world trade example network.
The unweighted version and the \nsi\ version using GDP as node weight plausibly 
place Europe and North Africa in one group, most of Asia in another,
and the Americas in the third, with little differences in the placement of some
``bordering'' countries.
(The placement of UGD (bottom left) in the blue group is an artifact of the algorithm 
due to its large distance from the network's centre.)
When using population as node weight, the \nsi\ result however differs considerably,
placing China and India in different groups, 
since the algorithm tends to produce groups of approximately equal weight.

\section{\label{sec:appl}%
  Application to climate networks
}

\noindent
Coming back to our original field of application, 
let us finally compare the unweighted and \nsi~versions of a number of network
measures in the case of a climate network whose node set is a
latitude-longitude-regular grid on the Earth's surface, with a resolution of
$2.5^\circ$ in latitude and $3.75^\circ$ in longitude. As in Donges et
al.\,\cite{Donges2009a,Donges2009b,Donges2011}, two of these $6,816$ nodes (the
two poles have been excluded) were linked when the corresponding time-series of
monthly averaged surface air temperature (SAT) anomalies from the 20th century
reference run 20c3m of the Hadley Centre's HadCM3 model (as defined in the
IPCC's Fourth Assessment Report, see~\cite{Donges2009a,Donges2009b} for details)
showed a significant Pearson correlation coefficient. We chose the link
inclusion threshold so as to achieve a relatively high link density of approx.\
0.05, for which all correlations of absolute value of at least 0.25 were
considered as significant.

\begin{figure}
  \includegraphics[width=\columnwidth]{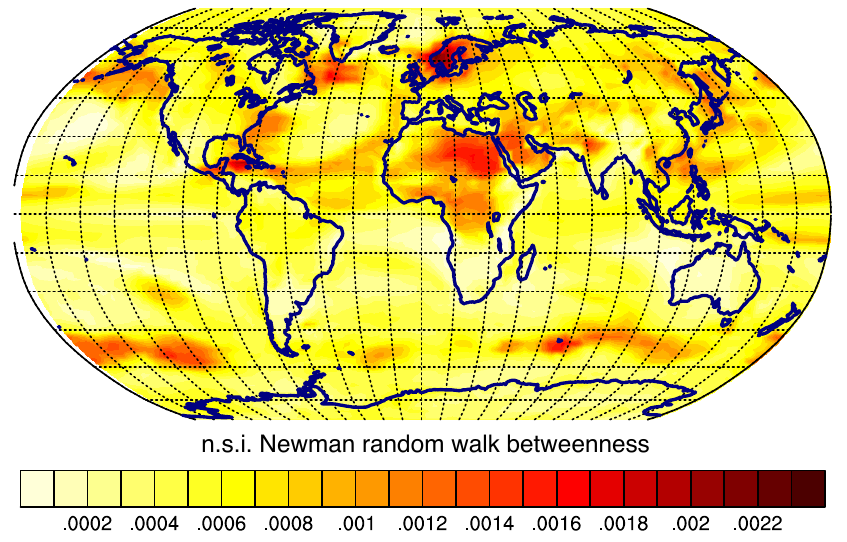}
  \caption[]{\label{fig:rwbglobe}
    (Colour online) \Nsi~version $NB^\ast_v$ of Newman's random walk betweenness in a
    global climate network representing correlations in surface air temperature dynamics 
    (same network as in Fig.\,\ref{fig:bylat}, Robinson projection).
    We can clearly identify the regions of the North Pacific Subpolar Gyre,
    the North Atlantic Subtropical Gyre including the Gulf Stream and the Canary Current,
    the North and South Equatorial Currents in the Pacific, and the Antarctic Circumpolar Current.
    The interpretation of other regions of high values like Scandinavia and Central and North-East Africa remains unclear.
  }
\end{figure}

As can be seen, e.\,g., from the resulting \nsi\ Newman-type random walk
betweenness (Fig.\,\ref{fig:rwbglobe}), this network retains a number of
interesting features of the global climate system. As argued in
\cite{Donges2009a}, increased values of Newman's random walk betweenness may be
indicative of diffusive transport processes in the climate system, e.\,g.,
turbulent eddy diffusion in the atmosphere and ocean, whereas shortest path
betweenness is believed to trace advective transport processes such as strong
surface ocean currents.

For comparison, we defined a second, synthetic network on the same set of nodes
in which we linked each pair of nodes $i,j$ independently with a probability of
$\min(1,\exp(0.4-0.09\alpha_{ij}))$, where $\alpha_{ij}$ is the angular distance
between the nodes (in degrees). The exponential relationship between link
probabity and distance was fitted to the relationship between the observed link
density and angular distance in the above climate network, using non-linear
regression (a similar relationship was found in \cite{Tsonis2008}). The
resulting {\em benchmark network} had a slightly smaller link density of
approx.\ 0.035 and can be interpreted as a sample from an underlying continuous
network whose link distribution depends on angular distance alone and is
therefore rotationally and translationally symmetric. Because of this underlying
symmetry, local network measures suitable for the estimation of underlying
features should not show a significant depency on the node's latitude. The thin
dashed and dotted lines in Fig.\,\ref{fig:bylat} show the longitudinally
averaged values of several network measures, plotted against latitude, in this
benchmark network. We can see that the \nsi~versions (dotted lines, using
cos(latitude) as node weight) fulfil this requirement of latitude independence
much better than the unweighted versions (thin dashed lines), which exhibit a
clear systematical increase either towards the poles (degree, clustering
coefficient, and closeness centrality) or towards the equator (Newman-type
random walk betweenness). For the degree, this is due to the increase in {\em
absolute} node density, while for the clustering coefficient, this is due to the
increase of the density {\em gradient} towards the pole.

As can be expected from this, the corresponding values in the real-world climate
network also show differences between the unweighted version (solid lines) and
the \nsi\ version (thick dashed lines) towards the poles or towards the equator.
Unweighted Newman-type random walk betweenness, e.\,g., is higher in the region
of the North and South Equatorial Currents at about $\pm 10^\circ$ latitude,
while its \nsi\ version is higher in the region of the North Atlantic
Subtropical Gyre between $+15^\circ$ and $+60^\circ$ latitude, although both
show well-defined features in both regions.

The influence of the increased node density in the high latitudes on unweighted
network statistics becomes even more evident when focussing on the Arctic region
north of $+60^\circ$ (Fig.\,\ref{fig:north}). While the unweighted degree and
clustering coefficient are markedly increased close to the North Pole, their
geographic distribution is considerably obscured by the node density induced
bias further southwards (Fig.\,\ref{fig:north}\,(A,\,B)). In contrast, the \nsi\
variants of degree and clustering coefficient reveal more pronounced regional
structures, e.\,g., increased \nsi\ degree over southern Greenland and
Scandinavia, or increased \nsi\ clustering coefficient surrounding Greenland
(Fig.\,\ref{fig:north}\,(C,\,D)). Hence, we judge \nsi\ network measures to be
very promising tools for future analysis of gridded climate data with
inhomogeneous mesh cell areas or station data, particularly since the additional
error that would be introduced by interpolation to equal area (geodesic) grids
can thus be avoided.

\section{\label{sec:conclusion}%
  Conclusion
}

\noindent
To summarize, in this article we have introduced a fairly general framework to deal
with biases and artifacts in complex network statistics that appear
when the nodes represent differently sized parts of an underlying domain of interest.
Networks of this type are routinely studied in various fields of research, 
including neuroscience, the Earth sciences, informatics and engineering, social sciences, economics, 
and dynamical systems, 
as our examples show,
and network design choices made when describing the domain of interest by a complex network 
can have considerable effects on the results of network statistics.

The central axiomatic notion of node splitting (or twin merging) invariance provides an elegant means to tackle the 
problem of how to use information on the size of (the part of the domain of interest represented by) 
individual nodes in a way that is robust against different choices of 
node selection, grid, meshing, parcellation, aggregation, or coarse-graining.
Our framework allowed us to derive
consistently weighted versions of a large representative set of commonly used statistical network measures 
that quantify different aspects of
networks in which links represent some kind of similarity or closeness.
Despite the diversity of these \nsi\ measures, 
a simple set of design rules (given in Sec.~\ref{subsec:pragmatic}) 
guided the introduction of node weights into their definitions.
The resulting formulas are in most cases computationally no more demanding than the original ones, 
and slightly modified versions of standard algorithms with the same complexity can usually be used to apply them in practice.
Also, since the construction of all our measures is based on the same first principles, 
they work better together and allow for an easier interpretation than alternative ad hoc approaches might.

Most importantly, 
\nsi\ measures reflect the features of the domain of interest more accurately than classical unweighted measures.
This was demonstrated in particular in the case of a synthetic and a real-world climate network,
for which it was possible to compare the results of \nsi\ measures on a non-homogeneous grid
with those based on a homogeneous (geodesic) grid, 
both showing exactly the same features 
and avoiding the artifacts that were produced by unweighted measures on the non-homogeneous grid. 
We have further illustrated the applicability and practical relevance of our axiomatic approach
by qualitatively showing their effects in a number of semi-realistic example networks.
In particular, we showed how in many of these examples, 
the judgement of which parts of the network are the most central or otherwise structurally important 
can change when node weights are used.

Our results also indicate that 
the topological properties of network representations of technical infrastructure such as the internet 
depend on whether the sometimes considerably varying size of the individual subsystems chosen as nodes in the network representation 
is taken into account or not,
and we conjecture that \nsi\ measures will prove highly relevant and beneficial 
for the consistent analysis of the vulnerability of distributed technical systems.
When analyzing how the connectivity of the internet decreases due to targeted attacks \cite{Albert2000},
the size of both the attacked and the affected autonomous systems should be of obvious interest.
A more thorough study of the AS network used for illustration here is of course beyond the scope of this methodological paper.
It would have to construct the links much more carefully by taking additional data sources into account, as described in \cite{Chen2002},
and verify that the number of CIDR prefixes is indeed a suitable size measure for autonomous systems.
We also leave for future research the general question of 
how one should choose node weights suitable for a given network and research question. 

\begin{acknowledgement}
  This work has been financially supported by the Federal Ministry for
  Education and Research (BMBF) via the Potsdam Research Cluster for
  Georisk Analysis, Environmental Change and Sustainability
  (PROGRESS), and the Leibniz association (project ECONS). JFD thanks the German National Academic Foundation for financial support.
  We acknowledge the modeling groups, the Program for Climate Model Diagnosis
  and Intercomparison (PCMDI) and the WCRP's Working Group on Coupled Modelling (WGCM)
  for their roles in making available the WCRP CMIP3 multi-model data set. 
  Support of this data set
  is provided by the Office of Science, U.\,S.~Department of Energy.
  Some graph theoretical calculations have been performed using the software
  package \texttt{igraph} \cite{Csardi2006}.
  The brain network data was kindly provided by S.~Achard, CNRS, Grenoble.  
\end{acknowledgement}

\bibliographystyle{epj}
\bibliography{networks_weighting_refs}

\clearpage

\section*{%
	Appendix A: Measures corrected for a typical weight
}

\noindent
Since we introduced \nsi\ measures $f^\ast$ as weighted versions of existing network measures $f$, 
it is natural to consider the case where all node weights $w_v$ are equal to some constant value $\omega$.
Usually, because of the construction mechanisms (i) and (ii) given in Sec.~\ref{subsec:pragmatic}, 
$f^\ast$ will not exactly equal $f$ in that case but rather be some (usually simple) transformation of it,
and only for $N\to\infty$, $f^\ast$ is often asymptotic to $f$,
e.\,g., $k^\ast_v=(k_v+1)\omega=\omega k_v + |o(k_v)|\sim k_v$.
When one wants to compare values of $f^\ast$ with those of $f$, 
this behaviour presents some difficulties, 
hence we will present here for most of the treated measures
a second, somewhat more complex, {\em corrected \nsi} version of $f^\ast$,
denoted by $f^{\ast\omega}$, which is also \nsi,
but which involves a parameter $\omega>0$ and has the property that
\begin{align}
  f^{\ast\omega}=f\quad\text{whenever}\quad w_v=\omega\text{~for all~}v\in\nodes,
\end{align}
so that $f^{\ast\omega}$ can be compared to $f$ more easily than $^{\ast}$.
The parameter $\omega$ is then called the {\em typical weight} 
and can be thought of as a kind of resolution or scale on which the analysis focuses.  
The question of how a suitable value for $\omega$ can be estimated from $G$ and $w$
if the weights are not all equal is addressed later, in the next subsection.

In many cases, the following construction mechanisms are helpful in addition to (i)--(iv) given in Sec.~\ref{subsec:pragmatic}
to derive $f^{\ast\omega}$ from $f^\ast$:
\begin{enumerate}
  \item[(v)]    Correct for the effect of (i) by replacing each occurrence of $w_i$ by $w_i/\omega$, and
  \item[(vi)]   correct for the effects of (ii) and (iii) by subtracting suitable terms (often constants) from sums over nodes.
\end{enumerate}
It will be convenient to express the corrections in terms of the corrected
version of $W$, which can be called the {\em corrected \nsi\ number of nodes,}
\begin{align}
	N^{\ast\omega} = W/\omega.
\end{align} 

~

\noindent
In the example of {\em degree,} we apply both (v) and (vi) and put
\begin{align}
  k^{\ast\omega}_v = \textstyle\sum_{i\in\nodes^+_v}w_i/\omega - 1 = k^\ast_v/\omega - 1
\end{align}
which obviously reduces to $k_v$ if $w_v\equiv\omega$.
In case of non-constant weights, it can happen that $k^{\ast\omega}_v$ turns out to be negative
for some $v$ if $\omega$ is chosen too large. 
The same effect can happen for corrected \nsi\ versions of other measures,
as will be obvious from their definitions presented below.
In general, negative values can be avoided by lowering $\omega$ or by replacing them by zero.

~

\noindent
For the {\em local clustering coefficient,} we have
\begin{align*}
  C^\ast_v ( w_i\equiv\omega ) = \textstyle\frac{C_v k_v(k_v-1) + 3k_v + 1}{(k_v+1)^2} = C_v + |O(\frac{1}{k_v})|.
\end{align*}
A {\em corrected \nsi\ local clustering coefficient} can only be defined for the case that $k^{\ast\omega}_v>1$,
\begin{align}
  C^{\ast\omega}_v
    = \frac{(N^{\ast\omega})^2\langle a^+_{vi}a^+_{ij}a^+_{jv}\rangle_{ij}^w
    		-3k^{\ast\omega}_v-1}{k^{\ast\omega}_v(k^{\ast\omega}_v-1)}
    \leqslant 1,
\end{align}
where, following (vi), we subtract $(3k^{\ast\omega}_v+1)$
since in $\langle a^+_{vi}a^+_{ij}a^+_{jv}\rangle_{ij}^w$, 
the nodes $i$ and $j$ can be equal or equal to $v$.
For $\omega\leqslant w_v/2$, one can prove that $C^{\ast\omega}_v\geqslant 0$.

~

\noindent
Also {\em corrected \nsi\ closeness centrality} measures can be derived via (v) and (vi):
\begin{align}
  CC_v^{\ast\omega} &= \frac{1}{1/CC_v^\ast - 1/N^{\ast\omega}} \leqslant 1,\nonumber\\
  CC'^{\ast\omega}_v &= CC'^\ast_v - 1/2N^{\ast\omega}\leqslant 1,\\
  CC''^{\ast\omega}_v &= CC''^\ast_v - 1/N^{\ast\omega}\leqslant 1.\nonumber
\end{align}

~

\noindent
In case of {\em (shortest path) betweenness centrality} $BC^\ast_v=\langle n^\ast_{ab}(v)/n^\ast_{ab}\rangle_{ab}^w$,
we only have to divide $n^\ast_{ab}$ and $n^\ast_{ab}(v)$
by a suitable power of $\omega$, according to (v).
Subtractions as in (vi) are unnecessary since we did not extend any sums to derive $BC^\ast_v$ from $BC_v$.
Hence the corrected \nsi\ versions are
\begin{align}
  & n^{\ast\omega}_{ab} = \omega^{1-d_{ab}}n^\ast_{ab},\quad
  n^{\ast\omega}_{ab}(v) = \omega^{2-d_{ab}}n^\ast_{ab}(v), \nonumber\\
  & BC^{\ast\omega}_v = \langle n^{\ast\omega}_{ab}(v)/n^{\ast\omega}_{ab}\rangle_{ab}^w
    = \omega BC^\ast_v \in [0,\omega/w_v].
\end{align}

~

\noindent
For {\em random walk-based measures,}
a corrected \nsi\ version of the transition probabilities $p_{ij}$
is only obvious in the pathological case in which $\omega\leqslant\min_v w_v$
and when no isolated nodes exist. We could then put
\begin{align}
  p^{\ast\omega}_{ij} = a^+_{ij}(w_j/\omega-\delta_{ij})/k^{\ast\omega}_i.
\end{align}
For larger, more realistic choices of typical weight, 
this definition of $p^{\ast\omega}_{ii}$ would result in negative values for many $i$
and could thus not be interpreted as a transition probability.
We will therefore not present corrected versions of random-walk based measures.

~

\noindent
The {\em aggregate statistics} we presented also allow for corrected \nsi\ versions:
For the {\em global clustering coefficient} it is
\begin{align}
  C^{\ast\omega} = \langle C^{\ast\omega}_v\rangle_v^w,
\end{align}
while for {\em transitivity} it is
\begin{align}
  T^{\ast\omega}
    = \frac{(N^{\ast\omega})^3\langle a_{vi}a_{ij}a_{jv}\rangle_{vij}
    		-N^{\ast\omega}\left(3\langle k^{\ast\omega}_v\rangle_v^w+1\right)}%
    	{(N^{\ast\omega})^3\langle a_{iv}a_{vj}\rangle_{vij}
    		-N^{\ast\omega}\left(3\langle k^{\ast\omega}_v\rangle_v^w+1\right)}
    \leqslant 1.
\end{align}
For the {\em link density,} 
{\em average (geodesic) path length,}
and {\em global efficiency,} it is just
\begin{align}\textstyle
  \varrho^{\ast\omega} = \varrho^\ast - \frac 1{N^{\ast\omega}},\quad
  L^{\ast\omega} = L^\ast - \frac 1{N^{\ast\omega}},\quad
  E^{\ast\omega} = E^\ast - \frac 1{N^{\ast\omega}}.
\end{align}

~

\noindent
For {\em spectral analysis,} the corrected \nsi\ versions of {\em adjacency} $\sf A$ and {\em Laplacian} $\sf \Lambda$
are
\begin{align}\textstyle
  {\sf A}^{\ast\omega}&=\textstyle\frac 1\omega {\sf A}^\ast-{\sf I},
  & {\sf\Lambda}^{\ast\omega} &= {\rm diag}(\vec k^{\ast\omega})-{\sf A}^{\ast\omega} 
  = \textstyle\frac 1\omega {\sf\Lambda}^\ast, \nonumber\\
  {\sf A}^{\ast\prime\omega}&=\textstyle\frac 1\omega {\sf A}^{\ast\prime}-{\sf I},~
  & {\sf\Lambda}^{\ast\prime\omega} &= {\rm diag}(\vec k^{\ast\omega})-{\sf A}^{\ast\prime\omega}
  = \textstyle\frac 1\omega {\sf\Lambda}^{\ast\prime}.
\end{align}
In particular, 
$EC^\ast_v(w_i\equiv\omega)=EC_v$,
hence \nsi\ eigenvector centrality needs no correction,
and also unweighted and \nsi\ spectral bisections are identical if $w_i\equiv\omega$.

\subsection*{%
  Estimation of typical weight
}

\noindent
The usefulness of our corrected versions of the \nsi\ measures depends on a suitable choice of the typical weight $\omega$.
In applications in which the underlying domain of interest does not provide a natural choice for $\omega$,
we can try to determine a suitable $\omega$ from the network itself.
Such an ``estimate'' $\tilde\omega$ of $\omega$ should ideally
(i) depend monotonically on the node weights $w_v$,
(ii) lie in $[\min_v w_v,\max_v w_v]$,
(iii) be \nsi~itself,
and (iv) be small enough so that measures such as $k^{\ast\omega}_v$ and $C^{\ast\omega}_v$ 
are defined and non-negative for all or at least almost all nodes.
In addition, it would be nice if 
(v) $\tilde\omega$ is statistically robust, i.\,e., cannot change unboundedly by only local changes to the network.

To fulfil (i), (ii), and (v), the most natural choice seems to be the median node weight.
With a small adjustment, also (iii) is satisfied:
Define the {\em \nsi\ twin-adjusted weight} of $v$ as
\begin{align}\textstyle
  w'^\ast_v = \sum_{i\in \nodes^+_v} w_i\delta_{\nodes^+_i\nodes^+_v} \in [w_v,k^\ast_v]
\end{align}
and let $\tilde\omega_I$ be the $w$-weighted median of $w'^\ast_v$.
This fulfils (i)--(iii) and (v), though not necessarily (iv).
Only when there are pathologically many twins, $\tilde\omega_I$ might exceed $\max_v w_v$.

A different approach is to address (iv) first by putting
\begin{align}\textstyle
  w''^\ast_v = \frac 3 4 k^\ast_v - \sqrt{\frac 9{16} {k^\ast_v}^2 - \frac 1 2 \Delta^\ast_v}
  \in (0,\frac 1 2 k^\ast_v].
\end{align}
Using $\omega=\min_v w''^\ast_v$ would then ensure that for all $v$, $k^{\ast\omega}_v\geqslant 1$
and both $C^{\ast\omega}_v$ and $C'^{\ast\omega}_v$ are defined and non-negative except maybe for those few $v$ where the above minimum is attained.
When $v$ is not isolated, $w''^\ast_v\geqslant\min_v w_v$, but it may easily exceed $\max_v w_v$.
Hence a good choice that fulfils (i)--(iv), though not (v), is
\begin{align}\textstyle
  \tilde\omega_{II} = \min(\tilde\omega_I,\min_v w''^\ast_v).
\end{align}
Finally, a trade-off between (iv) and (v) can be made by using in this definition not $\min_v w''^\ast_v$
but a small quantile, say the first percentile $P^w_1(w''^\ast_v)$, of the $w$-weighted distribution of $w''^\ast_v$:
\begin{align}\textstyle
  \tilde\omega_{III} = \min(\tilde\omega_I,P^w_1(w''^\ast_v)).
\end{align}

\section*{%
	Appendix B: Some additional measures
}

\noindent
The {\em average (nearest) neighbours' degree} of $v$,
\begin{align} k_{nn,v} = \textstyle\sum_{i\in \nodes_v}{k_i}/{k_v}, \end{align}
represents the average size of the region a point linked to $v$ is linked to.
Using the plug-in mechanism (iv) and the correction mechanism (v) and (vi),
we define its weighted versions,
the {\em \nsi\ average (nearest) neighbours' degree} and
the {\em corrected \nsi\ average (nearest) neighbours' degree}, as
\begin{align*}
  k^\ast_{nn,v} &= \textstyle\sum_{i\in \nodes^+_v}{w_i k^\ast_i}/{k^\ast_v},\\
  k^{\ast\omega}_{nn,v} &= \textstyle\sum_{i\in \nodes^+_v}{w_i k^{\ast\omega}_i}/{\omega k^{\ast\omega}_v}-1.
\end{align*}
These are \nsi\ since their ingredients $k^\ast_i$, $k^\ast_v$, $k^{\ast\omega}_i$, and $k^{\ast\omega}_v$ are.
In case of constant weights $w_i\equiv\omega$, we have $k^{\ast\omega}_{nn,v} = k_{nn,v}$,
whereas the uncorrected version is then
\begin{align*}
  k^\ast_{nn,v} ( w_i\equiv\omega ) &= \textstyle\frac{k_{nn,v}k_v+2k_v+1}{k_v+1}\omega \\
    &= (k_{nn,v} + 2)\omega + O(\textstyle\frac{1}{k_v}).
\end{align*}

~

\noindent
The Pearson product-moment correlation coefficient
between the degrees of the two end nodes of each link in $G$
is called {\em degree correlation} or {\em assortativity}
and can be computed from the degrees and average neighbours' degrees as follows:
\begin{align}
  r = \frac{
      \langle k_v^2 k_{nn,v}\rangle_v \langle k_v\rangle_v - (\langle k_v^2\rangle_v)^2
    }{
      \langle k_v^3\rangle_v \langle k_v\rangle_v - (\langle k_v^2\rangle_v)^2
    }.
\end{align}
A {\em [corrected] \nsi\ degree correlation} is most easily found using the plug-in mechanism (iv):
\begin{align*}
  r^\ast &= \frac{
      \langle {k^\ast_v}^2 k^\ast_{nn,v}\rangle_v^w \langle k^\ast_v\rangle_v^w - (\langle {k^\ast_v}^2\rangle_v^w)^2
    }{
      \langle {k^\ast_v}^3\rangle_v^w \langle k^\ast_v\rangle_v^w - (\langle {k^\ast_v}^2\rangle_v^w)^2
    },\\
  r^{\ast\omega} &= \frac{
      \langle {k^{\ast\omega}_v}^2 k^{\ast\omega}_{nn,v}\rangle_v^w \langle k^{\ast\omega}_v\rangle_v^w - (\langle {k^{\ast\omega}_v}^2\rangle_v^w)^2
    }{
      \langle {k^{\ast\omega}_v}^3\rangle_v^w \langle k^{\ast\omega}_v\rangle_v^w - (\langle {k^{\ast\omega}_v}^2\rangle_v^w)^2
    }.
\end{align*}
Note that $r^\ast$ is the common Pearson correlation coefficient between $k^\ast_i$ and $k^\ast_j$
in the probabilistic model in which the link or self-loop $i$---$j$ is drawn with relative probability $w_iw_j$.
The latter measures the number of links in $G_0$ between the regions represented by $i$ and $j$,
hence $r^\ast$ estimates the correlation coefficient
between the degrees of the two end points of all links in $G_0$.

~

\noindent
Soffer and V\'azquez \cite{Soffer2005} justify a version of the {\em clustering coefficient} $C_v$
which is partially adjusted for degree correlations:
\begin{align}
  C'_v
  = \frac{N^2\langle a_{vi}a_{ij}a_{jv}\rangle_{ij}}{\sum_{i\in \nodes_v}(\min(k_i,k_v)-1)}
  \in [C_v,1].
\end{align}
Again using all four mechanisms (i)--(iv), we get their \nsi\ versions:
\begin{align*}
  C'^\ast_v
    &= \frac{W^2\langle a^+_{vi}a^+_{ij}a^+_{jv}\rangle_{ij}^w}{\sum_{i\in \nodes^+_v}w_i\min(k^\ast_i,k^\ast_v)}
    \in [C^\ast_v,1]
    \quad\mbox{and}\\
  C'^{\ast\omega}_v
    &= \frac{(N^{\ast\omega})^2\langle a^+_{vi}a^+_{ij}a^+_{jv}\rangle_{ij}^w-3k^{\ast\omega}_v-1}%
  {\sum_{i\in \nodes^+_v}\frac{w_i}{\omega}\min(k^{\ast\omega}_i,k^{\ast\omega}_v)-2k^{\ast\omega}_v}
    \in [C^{\ast\omega}_v,1],
\end{align*}
where we define the latter only for the case where
$\sum_{i\in \nodes^+_v}\frac{w_i}{\omega}\min(k^{\ast\omega}_i,k^{\ast\omega}_v)>2k^{\ast\omega}_v$.

~

\noindent
Bonacich \cite{Bonacich1987} defined a measure of $v$'s {\em power centrality}
based on the idea that a node's ``power'' in a network should be the sum of
a linear function of each of its neighbours' powers:
\begin{align}
  PC_v
  = \textstyle\sum_{i\in\nodes}a_{vi}(\alpha + \beta PC_i)
\end{align}
for some parameters $\alpha,\beta>0$.
This implicit definition is solved by the power centrality vector
\begin{align} \vec{PC} = (PC_i)_{i\in\nodes} = \alpha({\sf I}-\beta {\sf A})^{-1}\vec k, \end{align}
where $\vec k=(k_i)_{i\in\nodes}$ is the degree vector,
assuming that ${\sf I}-\beta {\sf A}$ is invertible (which will be true for a general choice of $\beta$).
To find the \nsi\ version, we make the defining equation \nsi~and require
that $v$'s power is the weighted sum of the linear function of its neighbours' powers:
\begin{align} 
	PC^\ast_v = \textstyle\sum_{i\in\nodes} a^+_{vi}w_i (\alpha + \beta PC^\ast_i). 
\end{align}
Whenever ${\sf I}-\beta {\sf A}^+{\sf D}_w$ is invertible, the solution
\begin{align} 
	\vec{PC}^\ast = \alpha({\sf I}-\beta {\sf A}^\ast)^{-1}\vec k^\ast 
\end{align}
is then the sought {\em \nsi\ power centrality} vector,
where $\vec k^\ast$ is the \nsi\ degree vector,
$\vec k^\ast=(k^\ast_i)_{i\in\nodes}$.

Following (v) and (vi) again, the corrected \nsi\ equation is
\begin{align}
  PC^{\ast\omega}_v = \textstyle\sum_{i\in\nodes} a^+_{vi}\frac{w_i}{\omega}(\alpha + \beta PC^{\ast\omega}_i)
                      - (\alpha + \beta PC^{\ast\omega}_v),
\end{align}
and its solution is the {\em corrected \nsi\ power centrality} vector
\begin{align}
  \vec{PC}^{\ast\omega} =
  \textstyle\alpha\left({\sf I}-\beta{\sf A}^{\ast\omega}\right)^{-1}\vec k^{\ast\omega}.
\end{align}

~

\noindent
The {\em random walk centrality} of $v$ as defined by Noh and Rieger \cite{Noh2004}
measures how fast a walk starting at a random node reaches $v$:
\begin{align}\textstyle
  RWC_v = p_v\Big/\sum_{t=0}^\infty\big((P^t)_{vv}-p_v\big).
\end{align}
The obvious \nsi\ version of this is
\begin{align*}
  RWC^\ast_v &= \textstyle p^\ast_v\Big/\sum_{t=0}^\infty\big(((P^\ast)^t)_{vv}-p^\ast_v\big).
\end{align*}

~

\noindent
The {\em normal matrix} ${\sf T}={\rm diag}(\vec k)^{-1}{\sf A}$ is also often used as the basis of a centrality measure
and has the \nsi\ versions
\begin{align}
	{\sf T}^\ast={\rm diag}(\vec k^\ast)^{-1}{\sf A}^\ast,\quad
	{\sf T}^{\ast\omega}={\rm diag}(\vec k^{\ast\omega})^{-1}{\sf A}^{\ast\omega}
\end{align} 
which both retain ${\sf T}$'s property of having row sums equal to one.

\subsection*{\label{subsec:modularity}%
  Modularity
}

\noindent
Following Newman \cite{Newman2006}, the {\em modularity} of a partition $\cal P$ of $\nodes$
expresses how well the groups defined by $\cal P$ are internally connected and separated from each other.
It is defined as the observed within-group link density
minus its expected value given the observed degree distribution:
\begin{align}
  {\sf Q} = \frac{\langle \delta_{{\cal P}(i){\cal P}(j)}B_{ij}\rangle_{ij}}{\langle a_{ij}\rangle_{ij}}
  \quad\mbox{with}\quad
  B_{ij} = a_{ij} - \frac{k_i k_j}{N\langle k_v\rangle_v},
\end{align}
where ${\cal P}(i)$ is that set in $\cal P$ which contains $i$.
The (corrected) \nsi\ versions of these are:
\begin{align*}
  {\sf Q}^\ast &= \frac{\langle \delta_{{\cal P}(i){\cal P}(j)}B^+_{ij}\rangle_{ij}^w}{\langle a^+_{ij}\rangle_{ij}^w},
  & B^+_{ij} &= a^+_{ij} - \frac{k^\ast_i k^\ast_j}{W\langle k^\ast_v\rangle_v^w},\\
  {\sf Q}^{\ast\omega} &= 
  \frac{\langle \delta_{{\cal P}(i){\cal P}(j)}B^{+\omega}_{ij}\rangle_{ij}^w-1/N^{\ast\omega}}%
  {\langle a^+_{ij}\rangle_{ij}^w-1/N^{\ast\omega}},
  & B^{+\omega}_{ij} &= a^+_{ij} - \frac{k^{\ast\omega}_i k^{\ast\omega}_j}{N^{\ast\omega}\langle k^{\ast\omega}_v\rangle_v^w}.
\end{align*}
${\sf Q}^\ast$ estimates the within-group link density in $G_0$
minus its expected value given the degree distribution in $G_0$,
for the partition of $\nodes_0$ induced by the partition $\cal P$ of $\nodes$.

~

\noindent
Similar to the matrices ${\sf A}$, ${\sf\Lambda}$, and ${\sf T}$,
also the {\em spectrum of the modularity matrix} ${\sf B}=(B_{ij})_{ij}$ can be made
\nsi, using the matrices ${\sf B}^+=(B^+_{ij})_{ij}$ and ${\sf B}^{+\omega}=(B^{+\omega}_{ij})_{ij}$:
\begin{align*}
  {\sf B}^\ast &= {\sf D}_w^{1/2} {\sf B}^+{\sf D}_w^{1/2},
  & {\sf B}^{\ast\omega}&=\textstyle\frac 1\omega {\sf D}_w^{1/2} {\sf B}^{+\omega}{\sf D}_w^{1/2} - {\sf I}.
\end{align*}

~

\noindent
Given a subset $g\subseteq N$ of the nodes, the {\em generalized modularity matrix}
is the $|g|\times|g|$ matrix ${\sf B}^{(g)}=(B^{(g)}_{ij})_{ij}$ with
\begin{align}\textstyle
  B^{(g)}_{ij} = B_{ij} - \delta_{ij}\sum_{v\in g}B_{iv}.
\end{align}
Newman \cite{Newman2006} uses the signs in the eigenvector of its largest positive eigenvalue
in an efficient iterative {\em network dividing algorithm} similar to Fiedler's spectral bisection method,
in which a hierarchical clustering tree is constructed top-down by starting with $g_0=\nodes$, 
bisecting into two groups $g_1,g_2$ according to the eigenvector signs, 
and then repeating with each $g_i$ thus obtained.  
The following versions of ${\sf B}^{(g)}$ can be used to derive similar \nsi\ network divisions,
as exemplified in Fig.\,\ref{fig:trade}:
\begin{align*}
  B^{(g)\ast}_{ij} &= \textstyle B^\ast_{ij} - \delta_{ij}\sum_{v\in g} w_v B^+_{iv},\\
  B^{(g)\ast\omega}_{ij} &= \textstyle B^{\ast\omega}_{ij}
    - \delta_{ij}\sum_{v\in g}(\frac{w_v}{\omega} B^{+\omega}_{iv} - \delta_{iv}).
\end{align*}
If we assume that when splitting $s\to s'+s''$ with $s\in g$, both $s'$ and $s''$ are put into the new $g$,
then the eigenvectors of these matrices have the same \nsi\ properties as those of ${\sf B}^\ast$ above.
In particular, the eigenvector entries for $s'$ and $s''$ have the same sign as that of $s$,
whereas all other signs are unchanged,
hence the division of $g$ will be the same after the split
except that $s$ is replaced by $s'$ and $s''$ in its subgroup.

\end{document}